\documentclass[twocolumn]{aastex62}

\received{February 12, 2019}
\revised{May 13, 2019}
\accepted{May 25, 2019}
\submitjournal{ApJ}

\shorttitle{Molecular Gas Density toward a Spiral Arm of M51}
\shortauthors{Nishimura et al.}

\begin{document}

\title{Molecular Gas Density Measured with H$_2$CO and CS 
       toward a Spiral Arm of M51}

\correspondingauthor{Yuri Nishimura}
\email{yuri@ioa.s.u-tokyo.ac.jp}

\author[0000-0003-0563-067X]{Yuri Nishimura}
\affiliation{Institute of Astronomy, 
The University of Tokyo, 
2-21-1, Osawa, Mitaka, Tokyo 181-0015, Japan}
\affiliation{Chile Observatory, 
National Astronomical Observatory of Japan, 
2-21-1, Osawa, Mitaka, Tokyo 181-8588, Japan}

\author{Yoshimasa Watanabe}
\affiliation{Faculty of Pure and Applied Sciences, 
University of Tsukuba, Tsukuba, Ibaraki 305-8577, Japan}
\affiliation{Tomonaga Center for the History of the Universe, 
University of Tsukuba, Tsukuba, Ibaraki 305-8571, Japan}

\author{Nanase Harada}
\affiliation{Institute of Astronomy and Astrophysics, Academia Sinica, 
11F of AS/NTUAstronomy-Mathematics Building, No.1, Sec. 4, 
Roosevelt Rd, Taipei 10617, Taiwan, R.O.C.}

\author{Kotaro Kohno}
\affiliation{Institute of Astronomy, 
The University of Tokyo, 
2-21-1, Osawa, Mitaka, Tokyo 181-0015, Japan}
\affiliation{Research Center for the Early Universe, 
The University of Tokyo, 
7-3-1 Hongo, Bunkyo, Tokyo 113-0033}

\author{Satoshi Yamamoto}
\affiliation{Department of Physics, 
The University of Tokyo, 
7-3-1 Hongo, Bunkyo, Tokyo 113-0033}
\affiliation{Research Center for the Early Universe, 
The University of Tokyo, 
7-3-1 Hongo, Bunkyo, Tokyo 113-0033}

\begin{abstract}
Observations of various molecular lines toward a disk region of 
a nearby galaxy are now feasible, and they are being employed as 
diagnostic tools to study star-formation activities there. 
However, the spatial resolution attainable for a nearby galaxy 
with currently available radio telescopes is $10-1000$ pc, 
which is much larger than the scales of individual star-forming 
regions and molecular cloud cores. Hence, it is of fundamental 
importance to elucidate which part of an interstellar medium 
such spatially-unresolved observations are tracing. 
Here we present sensitive measurements of the H$_2$CO ($1_{01}-0_{00}$) 
line at 72 GHz toward giant molecular clouds (GMCs) in the spiral 
arm of M51 using the NRO 45 m and IRAM 30 m telescopes. 
In conjunction with the previously observed H$_2$CO ($2_{02}-1_{01}$) 
and CS ($2-1$ and $3-2$) lines, we derive the H$_2$ density 
of the emitting regions to be $(0.6-2.6)\times10^4$~cm$^{-3}$ 
and $(2.9-12)\times10^4$~cm$^{-3}$ for H$_2$CO and CS, respectively, 
by the non-LTE analyses, where we assume the source size of $0.8-1$ kpc 
and the gas kinetic temperature of $10-20$ K. The derived H$_2$ density 
indicates that the emission of H$_2$CO and CS is not localized to 
star-forming cores, but is likely distributed over an entire region of GMCs. 
Such widespread distributions of H$_2$CO and CS are also supported 
by models assuming lognormal density distributions over the 1 kpc region. 
Thus, contributions from the widespread less-dense components should 
be taken into account for interpretation of these molecular emission 
observed with a GMC-scale resolution. The different H$_2$ 
densities derived for H$_2$CO and CS imply their different distributions. 
We discuss this differences in terms of the formation processes of H$_2$CO and CS.
\end{abstract}


\section{Introduction} \label{sec:intro}

Recently, molecular line observations in the millimeter and 
submillimeter wave regimes have become more and more popular in 
extragalactic studies \citep[e.g.,][]{Aladro2015,Meier2015,Harada2018}. 
Thanks to the advanced instrumental capabilities of single-dish 
telescopes and interferometers, rotational spectral lines of various 
molecular species other than CO and its isotopologues can readily 
be detected in external galaxies. For instance, HCN and HCO$^+$, 
which are often employed as dense-gas ($> 3\times10^{4}$ cm$^{-3}$) tracers, 
have extensively been observed to reveal physical conditions of a molecular gas 
\citep[e.g.,][]{Gao2004,Usero2015,Bigiel2016,Jimenez-Donaire2017}. 
H$_2$CO and CS are also popularly observed to measure the gas density 
and temperature \citep[e.g.,][]{Mauersberger1989,Bayet2009,Mangum2013,Tang2017}.
Furthermore, molecular inventory is now being extended not only 
to bright central regions of galaxies but also to faint disk regions 
\citep[e.g.,][]{Watanabe2014,Bigiel2016,Watanabe2019}. However, 
the spatial resolution that can be achieved with currently available 
radio telescopes is $10-1000$ pc even for nearby galaxies: an exception is 
the nearest galaxies such as the Large and Small Magellanic Clouds. 
This resolution is larger than typical sizes of star forming regions 
($\sim0.1$ pc) and giant molecular clouds (GMCs, $\sim1-10$ pc). 
To make a full use of molecular lines for physical and chemical 
diagnostics of disk regions in external galaxies, we need to know 
what parts of an interstellar medium the emission of the dense-gas 
tracers comes from. More specifically, it is important to validate 
whether the emission of each molecular species is exclusively localized to 
cloud cores or distributed also in the diffuse regime.

There are two approaches to this problem. One approach is to conduct 
a large-scale mapping observation toward Galactic GMCs with various 
molecular lines, and to evaluate a fraction of the emission coming 
from each part of GMCs (i.e., star forming cores, their envelopes, 
and cloud peripheries) for each molecular line when the line is observed 
with a large beam covering a whole GMCs. This approach is being conducted in 
several GMCs \citep{Harada2019,Kauffmann2017,Nishimura2017,Pety2017,Watanabe2017}. 
These studies indicate a significant contribution of the emission 
from less dense regions even for ``dense-gas'' tracers such as HCN, 
HCO$^+$, and CS. The other approach, which is directly 
applicable for extragalactic sources, is to measure the gas density of 
the emitting region by multi-line observations. So far, multi-line 
observations have been conducted toward central regions of galaxies 
\citep[e.g.,][]{Aladro2011a}. On the other hand, such observations 
are very challenging toward disk regions because molecular lines are 
fainter \citep[e.g.,][]{Usero2015,Bigiel2016}. Consequently, 
the gas density traced by each molecular line is not fully understood 
particularly in disk regions. Nonetheless, the gas density in relatively 
quiescent disk regions is crucial for total understanding of star formation 
activities in a whole galaxy. 

With this in mind, we have conducted very sensitive observations 
of the lowest transition line of H$_2$CO ($1_{01}-0_{00}$) toward 
a position in a spiral arm of the nearby spiral galaxy M51 
\citep[$D=8.4$ Mpc;][]{Vinko2012}. In conjunction with the previously 
observed $2_{02}-1_{01}$ line \citep{Watanabe2014}, we evaluate 
the gas density of the emitting region of H$_2$CO by non-local 
thermal equilibrium (LTE) analyses. The observation of the 
$1_{01}-0_{00}$ is essential to derive the gas density in a cold and 
less dense conditions. In addition, we derive the gas density of 
the emitting region of CS by using the existing data of the two CS lines 
\citep[$J=2-1$ and $3-2$;][]{Watanabe2014}. Based on the derived 
quantities, we consider the realistic H$_2$ density distribution within 
the telescope beam ($\sim 1$ kpc). We also present implication of 
the derived results to formation processes of H$_2$CO and CS. 

\section{Observations and results} \label{sec:obs}

\begin{figure}
\centering
\includegraphics[width=0.9\hsize]{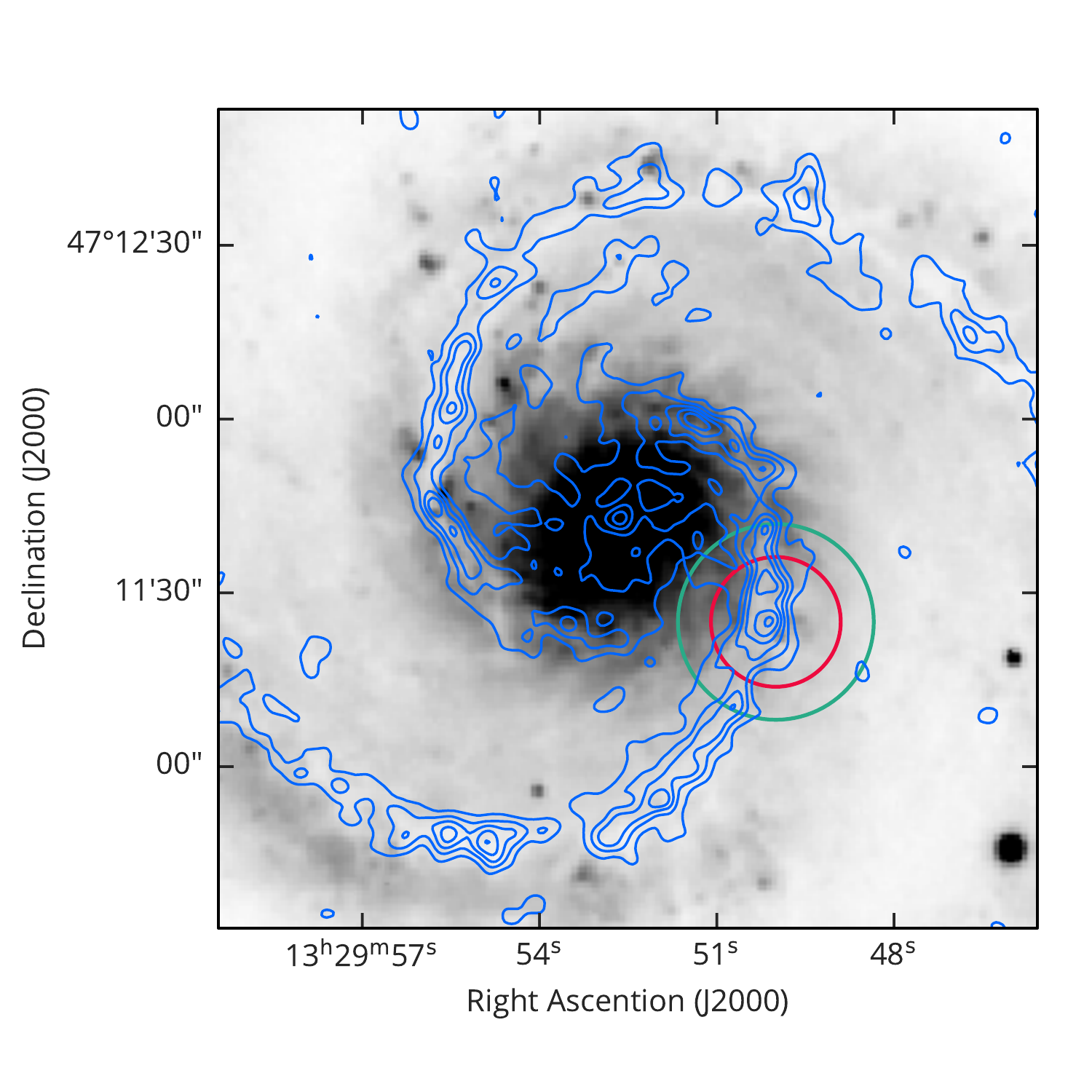}
\caption{The observed position M51 P1 and the beam size of NRO 45 m 
({22.4\arcsec}, red circle) and IRAM 30 m ({33.8\arcsec}, green circle)
at the observing frequency of the H$_2$CO ($1_{01}-0_{00}$) line (72.72 GHz) 
overlaid on the \textit{R}-band image (gray scale) by \citet{Hoopes2001} 
and the $^{12}$CO ($1-0$) integrated intensity (blue contours) 
by PAWS \citep{Schinnerer2013}. 
The contour levels are from 40 to 240 K km s$^{-1}$ 
at an interval of 40 K km s$^{-1}$. }
\label{fig:position}
\end{figure}

\begin{figure*}
\centering
\includegraphics[width=0.45\hsize]{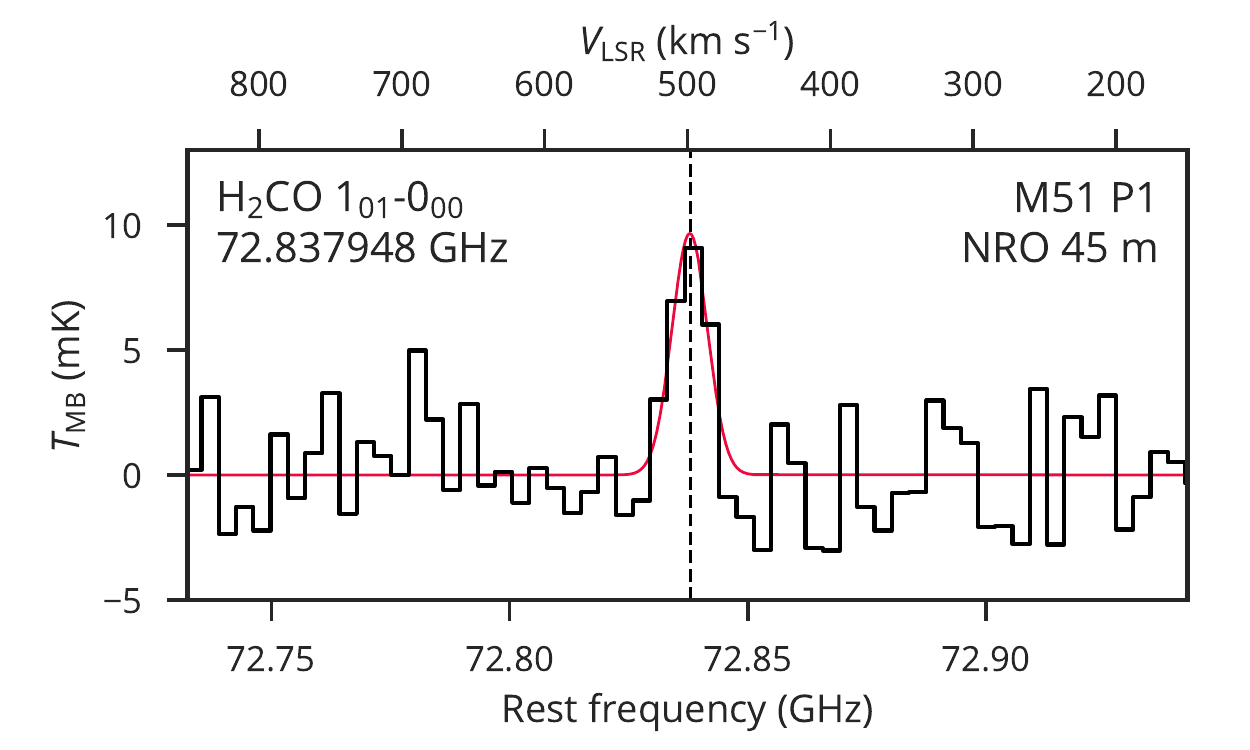}\hspace{0.03\hsize}
\includegraphics[width=0.45\hsize]{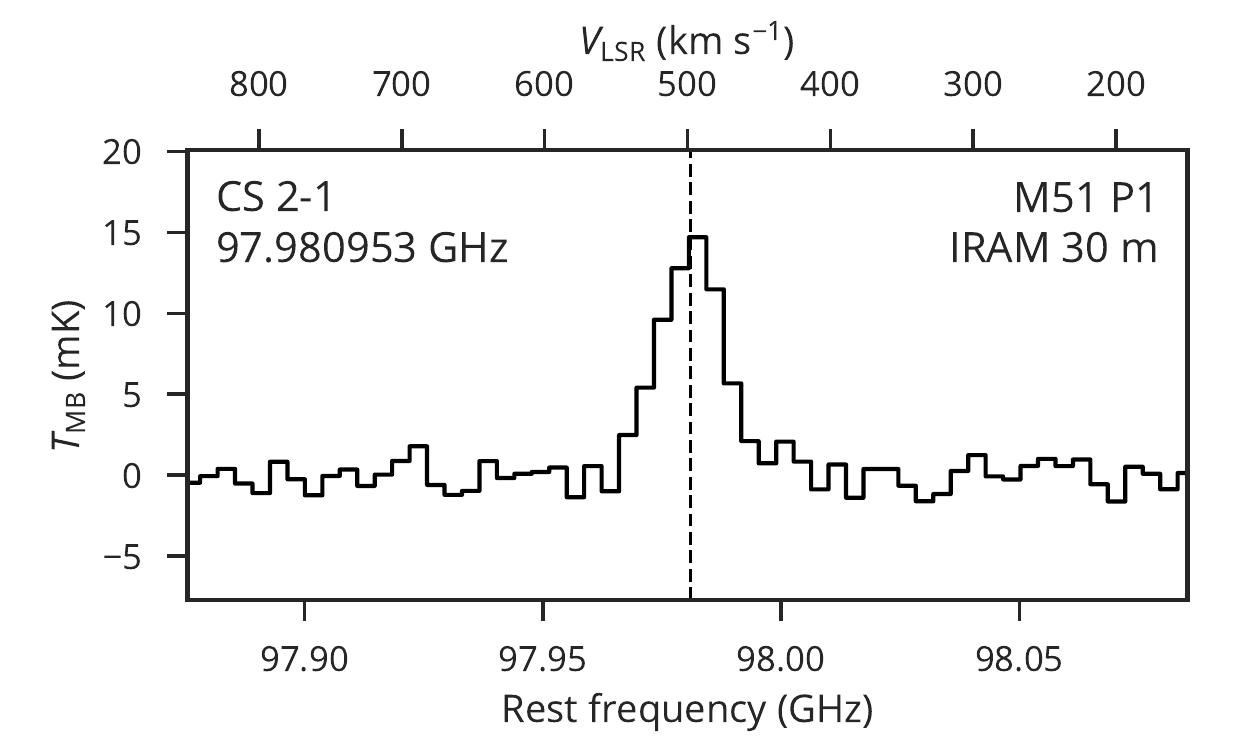}\\
\includegraphics[width=0.45\hsize]{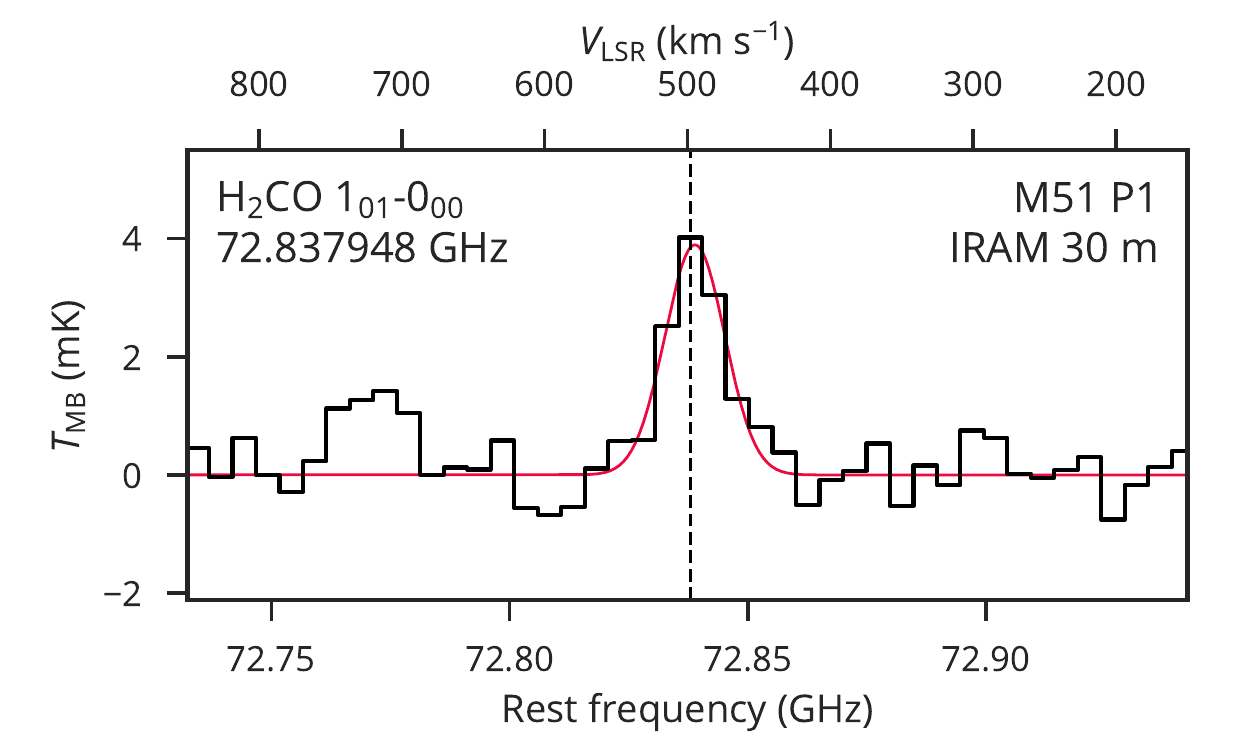}\hspace{0.03\hsize}
\includegraphics[width=0.45\hsize]{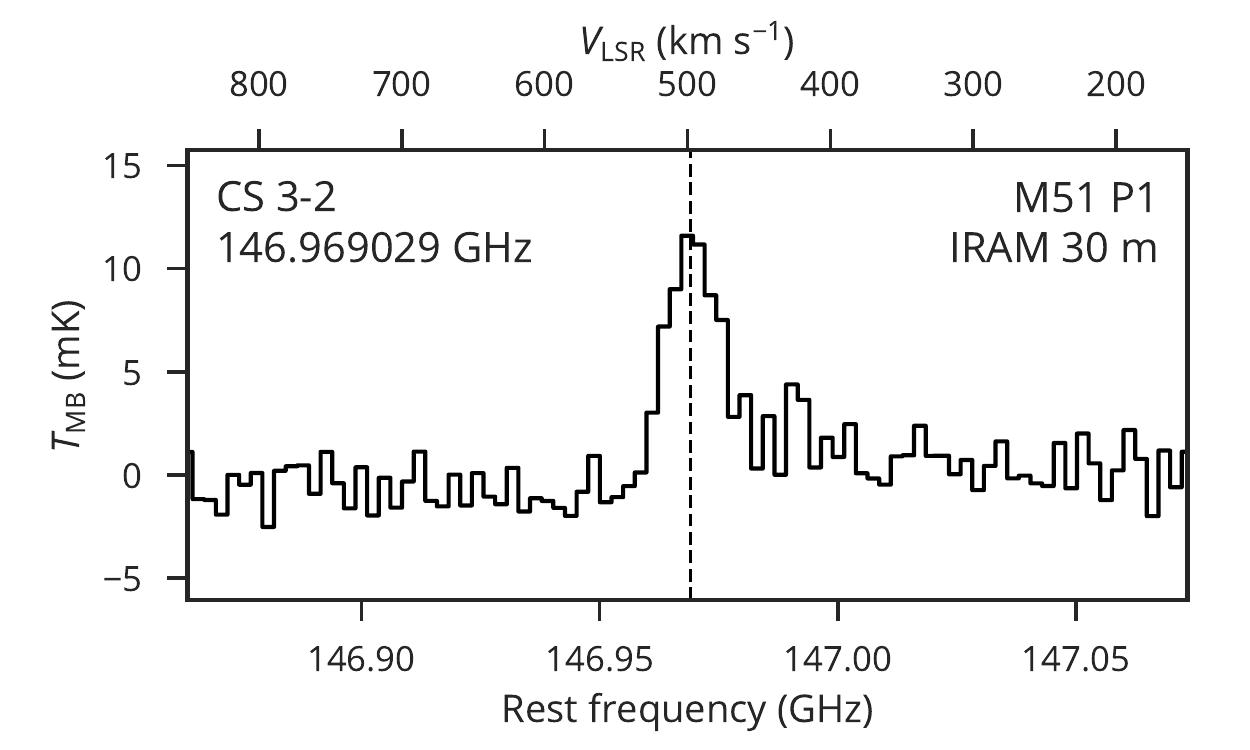}\\
\includegraphics[width=0.45\hsize]{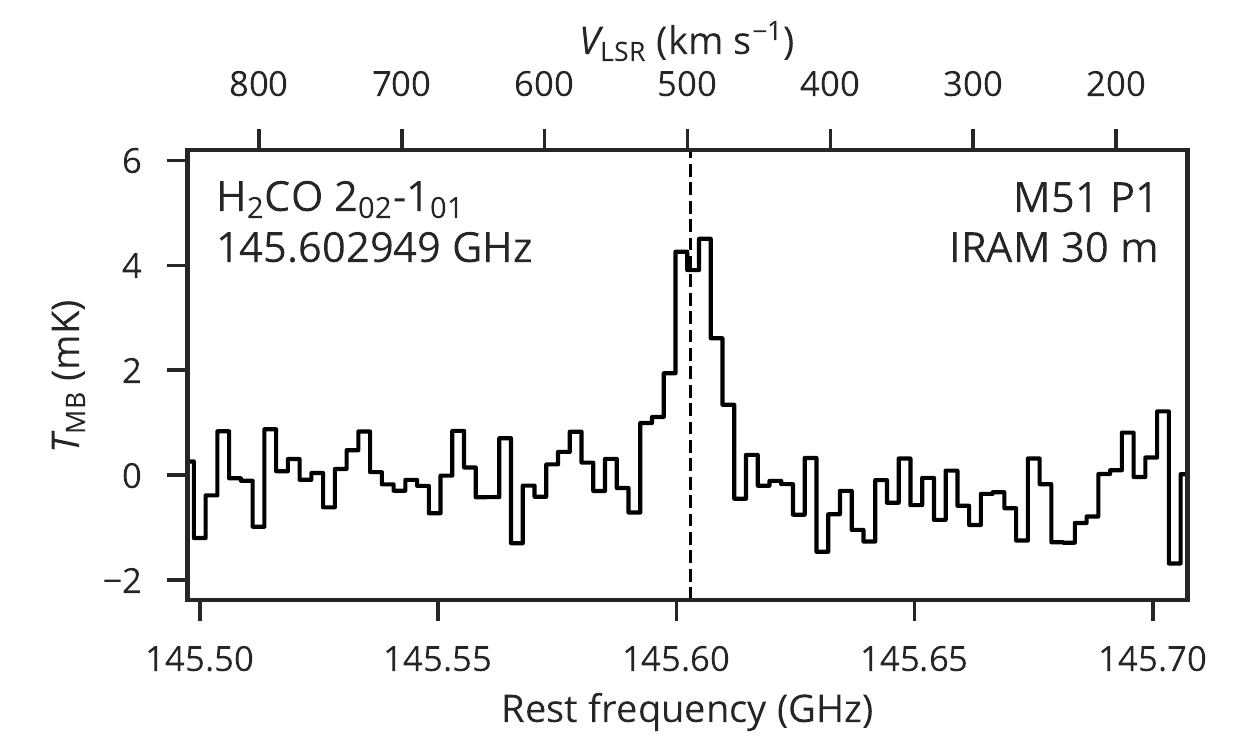}\hspace{0.03\hsize}
\includegraphics[width=0.45\hsize]{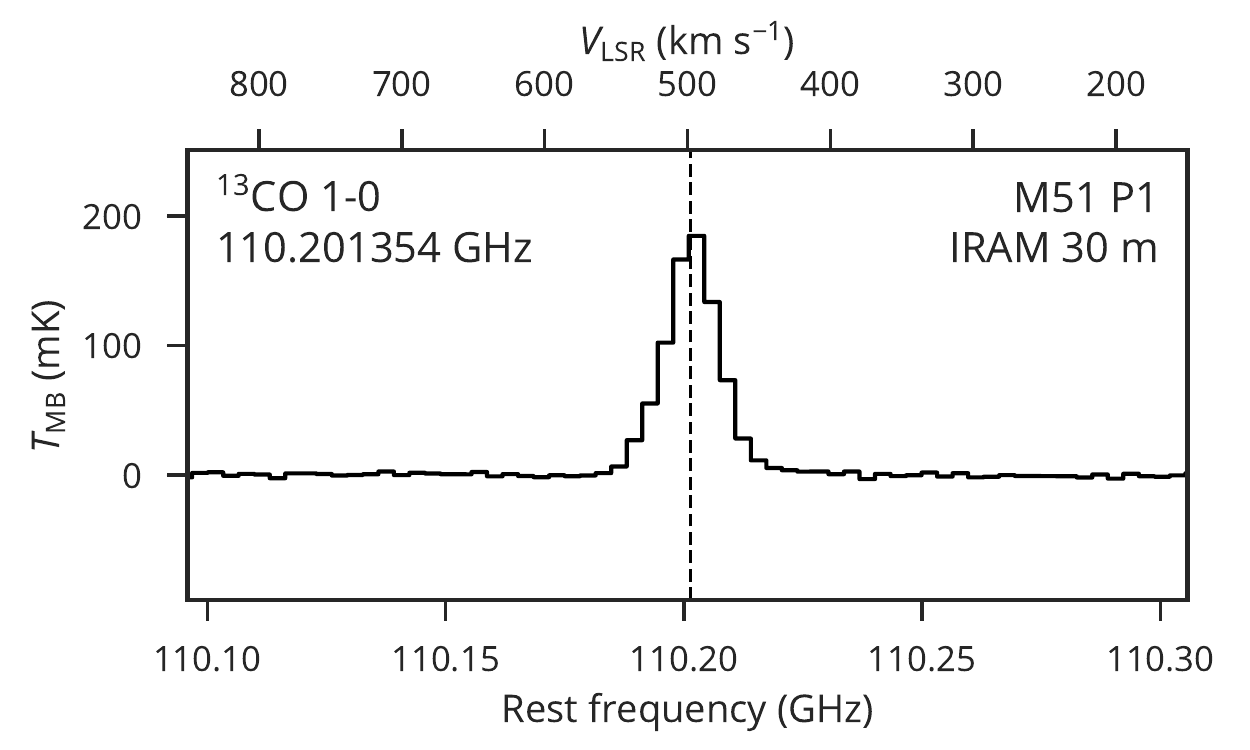}\\
\caption{The H$_2$CO ($1_{01}-0_{00}$) spectrum at 72.83798 GHz toward 
M51 P1 obtained with NRO 45 m (\emph{top left}) and IRAM 30 m 
(\emph{middle left}). The red curves show the results of a single Gaussian fit. 
The spectra of H$_2$CO ($2_{02}-1_{01}$; \emph{bottom left}), 
CS ($2-1$; \emph{top right}), CS ($3-2$; \emph{middle right}), and 
$^{13}$CO ($1-0$; \emph{bottom right}) obtained by \citet{Watanabe2014} 
are also shown for comparison. The dashed vertical lines indicate 
the LSR velocity of 498 km s$^{-1}$. The rest frequency is represented 
by assuming the LSR velocity of 498 km s$^{-1}$ for all the panels. 
Note that the temperature scale of the spectrum is different from panel 
to panel.}
\label{fig:lineprofile}
\end{figure*}

Observations of the H$_2$CO ($1_{01}-0_{00}$) line at 
72.8379480 GHz were carried out with the 45 m radio telescope 
at the Nobeyama Radio Observatory (NRO 45 m) in 2014 May. 
The half power beam width is $\sim22\farcs4$ at the observing 
frequency (72.72 GHz). We used the dual-polarization side-band 
separating (2SB) receiver T70. 
The system temperature ranged from 180 to 270 K. 
The sideband separation was typically $10-15$ dB or better. 
The backend was the autocorrelator SAM45 \citep{Kuno2011, Kamazaki2012}. 
The frequency resolution and bandwidth are 488.28 kHz and 1600 MHz, 
respectively. The antenna temperature $T_\mathrm{a}^*$ was corrected for 
the main beam efficiency of 0.45 to obtain the main beam temperature 
$T_\mathrm{MB}$\footnote{%
\url{http://www.nro.nao.ac.jp/~nro45mrt/html/prop/eff/eff2013.html}}. 
We employed the position-switching mode, where the on-source 
integration time of each scan was set to be 20 seconds. 
The observed position was M51 P1 
($\alpha_\mathrm{J2000}$ = 13$^\mathrm{h}$ 29$^\mathrm{m}$ {50\fs0}, 
$\delta_\mathrm{J2000}$ = {+47\arcdeg} {11\arcmin} {25\farcs0}), 
toward which \citet{Watanabe2014} conducted a spectral line survey 
in the 3 mm and 2 mm bands using the IRAM 30 m (Figure \ref{fig:position}). 
The position is the brightest $^{12}$CO ($1-0$) peak in the spiral arm. 
It contains H$\alpha$ and Pa$\alpha$ emission spots and is also bright 
in 24 $\mu$m continuum emission, all of which indicate that star 
formation occurs inside \citep{Schinnerer2010,Egusa2011}. 
The off-source position was {10\arcmin} away in azimuth from the 
on-source position. The telescope pointing was checked every 1--1.5 hours 
by observing nearby SiO maser sources (S-UMi and R-Cvn). 
The on-source integration time and the total observation time were 
18 hr and 50 hr, respectively. The observation data were reduced with 
the NRO software \emph{NEWSTAR}. In the analysis, we binned 
six successive channels of SAM45 to improve the signal-to-noise ratio. 
The resultant velocity resolution is 12.2 km s$^{-1}$ at 72 GHz. 
The rms noise temperature was 2 mK in the $T_\mathrm{MB}$ scale. 
Then, a baseline of the 5th-order polynomial was subtracted in the 
velocity range from $-200$ to $1200$ km s$^{-1}$. The line parameters 
were obtained by a single Gaussian fitting (Figure \ref{fig:lineprofile} 
top left), as summarized in Table \ref{tab:lineparameter}. 

The H$_2$CO ($1_{01}-0_{00}$) line was also observed with the IRAM 
30 m telescope in 2018 January and May, as a part of 
the 70 GHz-band line survey toward the corresponding position in M51 
(Y. Watanabe et al.~in preparation). The observation was 
conducted by using the dual-sideband dual-polarization EMIR receiver%
\footnote{\url{http://www.iram.es/IRAMES/mainWiki/EmirforAstronomers}} 
and the Fourier transform spectrometers. The half power beam width of 
IRAM 30 m at the frequency of the H$_2$CO ($1_{01}-0_{00}$) line is 
33\farcs8. We performed a single Gaussian fitting to the observed line 
profile (Figure \ref{fig:lineprofile} middle left) and obtained the 
line parameters, as shown in Table \ref{tab:lineparameter}. 

In addition to the newly observed data, we use the existing data of 
the H$_2$CO ($2_{02}-1_{01}$) and CS ($2-1$ and $3-2$) lines observed 
with IRAM 30 m \citep{Watanabe2014}. For these lines, we employ 
the line parameters reported by \citet{Watanabe2014}. 

\begin{deluxetable*}{lccccccccc}
\tabletypesize{\scriptsize}
\tablecaption{Observed Line Parameters \label{tab:lineparameter}}
\tablehead{
\colhead{Molecule} & \colhead{Transition} & \colhead{Frequency} 
 & \colhead{$E_\mathrm{up}$} & \colhead{$\int T_\mathrm{MB} dv$} 
 & \colhead{FWHM} & \colhead{$v_\mathrm{LSR}$} & \colhead{$T_\mathrm{MB}$ peak}
 & \colhead{$\theta_\mathrm{beam}$} & \colhead{Reference} \\
\colhead{} & \colhead{} & \colhead{(GHz)}
 & \colhead{(K)}& \colhead{(K km s$^{-1}$)} 
 & \colhead{(km s$^{-1}$)} & \colhead{(km s$^{-1}$)} & \colhead{(mK)} 
 & \colhead{(\arcsec)} & \colhead{}
}
\startdata
H$_2$CO & $1_{01}-0_{00}$ & 72.837948 & 3.5 
 & $0.30\pm0.09$ & $29\pm19$ & $498\pm3$ & $10\pm6$ 
 & 22.4 & This work \\
 & & & & $0.19\pm0.03$ & $49\pm17$ & $495\pm7$ & $4\pm1$ 
 & 33.8\tablenotemark{a} 
 & Watanabe et al.~(in prep.) \\
H$_2$CO & $2_{02}-1_{01}$ & 145.602949 & 10.5 
 & $0.14\pm0.03$ & $34\pm3$ & $495\pm1$ & $4\pm1$ 
 & 16.9\tablenotemark{a} & \citet{Watanabe2014} \\
CS & $2-1$ & 97.980953 & 7.1 
 & $0.68\pm0.08$ & $52\pm3$ & $497\pm1$ & $13\pm2$ 
 & 25.1\tablenotemark{a} & \citet{Watanabe2014} \\
CS & $3-2$ & 146.969029 & 14.1 
 & $0.4\pm0.1$ & $45\pm4$ & $495\pm2$ & $9\pm3$ 
 & 16.7\tablenotemark{a} & \citet{Watanabe2014}
\enddata
\tablenotetext{a}{The half power beam width of IRAM 30 m at the given 
frequency $\nu$ is 2460\arcsec $\cdot$($\nu$/GHz)$^{-1}$.}
\tablecomments{The uncertainties are 3$\sigma$.}
\end{deluxetable*}

\section{Non-LTE analyses} \label{sec:analyses}

\subsection{Non-LTE analyses using RADEX} \label{subsec:RADEX}

In order to derive the gas density, we conduct non-LTE analyses of 
the observed data. We use the publicly available code RADEX 
\citep{vanderTak2007} with collisional rate coefficients
of H$_2$CO \citep{Wiesenfeld2013} and CS \citep{Lique2006}. 
RADEX requires five input parameters to calculate the intensities of 
molecular lines: background temperature, line width, column density 
of a given species, gas kinetic temperature, and H$_2$ density. 
We select the H$_2$ density and the column density as adjustable 
parameters to reproduce the observed integrated intensities. 

Prior to the analyses, the intensity of each molecular line 
is corrected for the beam dilution effect. Beam dilution is caused 
by the coupling between the source and the telescope beam, as 
$T_\mathrm{B} = [({\theta_\mathrm{s}}^2+{\theta_\mathrm{b}}^2)
/{\theta_\mathrm{s}}^2] T_\mathrm{MB}$, where $T_\mathrm{B}$ is 
the source-averaged brightness temperature, $\theta_\mathrm{s}$ is 
the source size, $\theta_\mathrm{b}$ is the beam size of the telescope, 
and $T_\mathrm{MB}$ is the measured main-beam temperature. For H$_2$CO, 
the source size is derived to be {25\arcsec} (1 kpc) by using the two 
measurements of the H$_2$CO ($1_{01}-0_{00}$) line with the different 
telescope beams (22\farcs4 and 33\farcs8). Such a wide distribution seems 
likely, because H$_2$CO is ubiquitously detected toward Galactic clouds 
\citep[e.g., $\sim80$\% of 262 Galactic radio sources;][]{Downes1980}. 
For CS, we assume the source size on the basis of the interferometric 
observations in the literature: \citet{Watanabe2016} mapped the same 
region in the CS ($2-1$) line with the Combined Array for Research in 
Millimeter-wave Astronomy (CARMA) at a {$\sim6$\arcsec} resolution. 
The CS emission has a distribution extending along the spiral arm 
with a size of {$\sim20$\arcsec} (0.8 kpc). 
The flux resolved out by the interferometer (missing flux) is 
estimated to be $\sim30\%$ by comparing with the flux obtained 
with IRAM 30 m \citep{Watanabe2014}. 
We should keep in mind that the flux from the extended component 
may contribute appreciably in the single dish observations. 
Hence, we assume the two source sizes of {25\arcsec} and {20\arcsec} 
for both H$_2$CO and CS in the following analyses. 

The assumptions in the analyses are as follows: 
we adopt the cosmic microwave background temperature of 2.73 K. 
For the line width, we use 37 km s$^{-1}$ for H$_2$CO and 48 km s$^{-1}$ for CS, 
taking an average of the observed values (Table \ref{tab:lineparameter}). 
As for the gas kinetic temperature, \citet{Schinnerer2010} 
derive it to be $16-20$ K by a large velocity gradient (LVG) modeling of 
the line intensity ratios of $^{12}$CO ($1-0$), $^{12}$CO ($2-1$) and 
$^{13}$CO ($1-0$). Hence, we assume a somewhat wide range of the 
temperature, i.e., 10, 15, and 20 K. We run the RADEX code with the 
above parameters, where model grids of the H$_2$ density and the column 
density are set to 30 values logarithmically spaced from $10^2$ to 
$10^6$ cm$^{-3}$ and 30 values logarithmically spaced from $10^{10}$ 
to $10^{15}$ cm$^{-2}$, respectively. 

\begin{figure*}
\centering
\includegraphics[width=0.45\hsize]{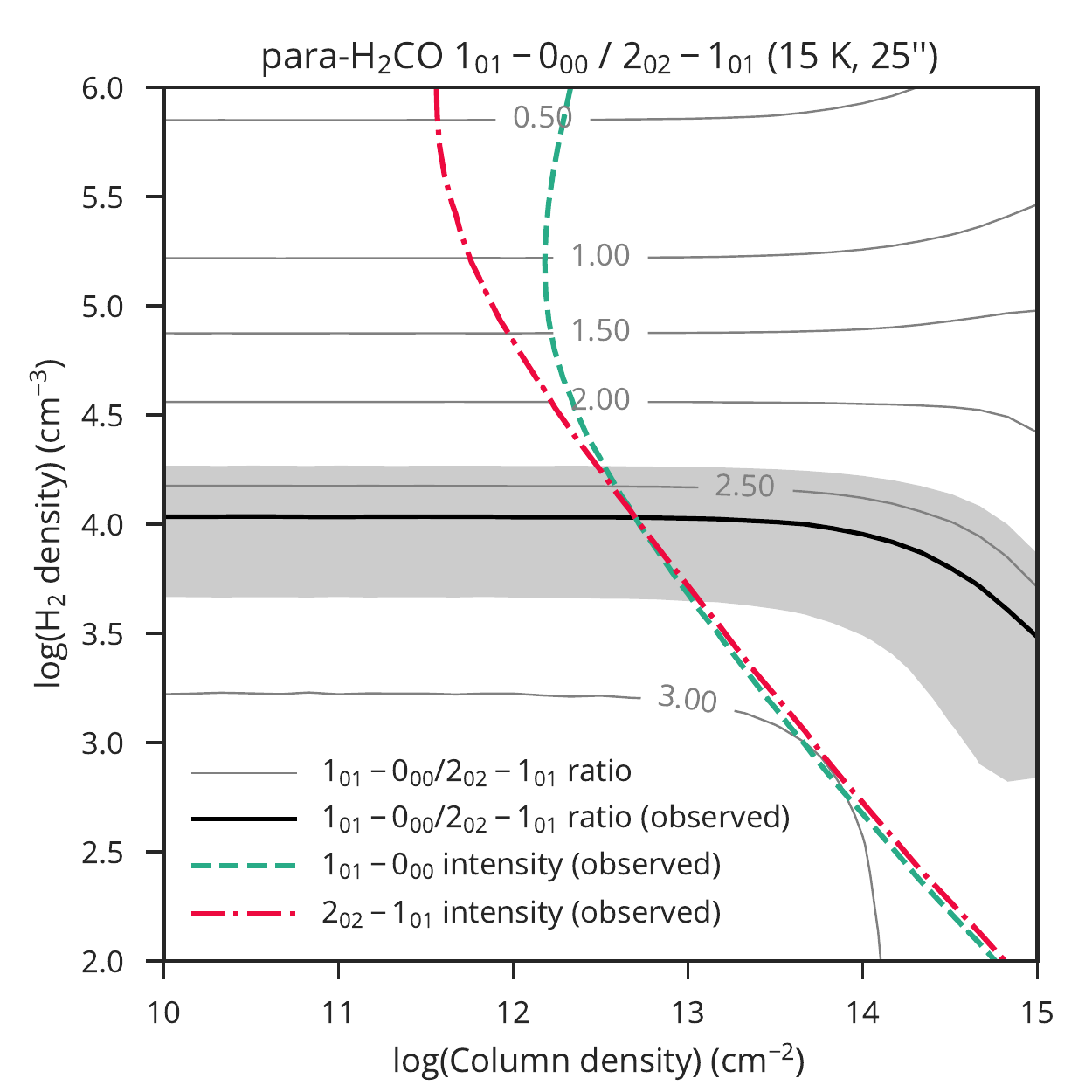}\hspace{0.03\hsize}
\includegraphics[width=0.45\hsize]{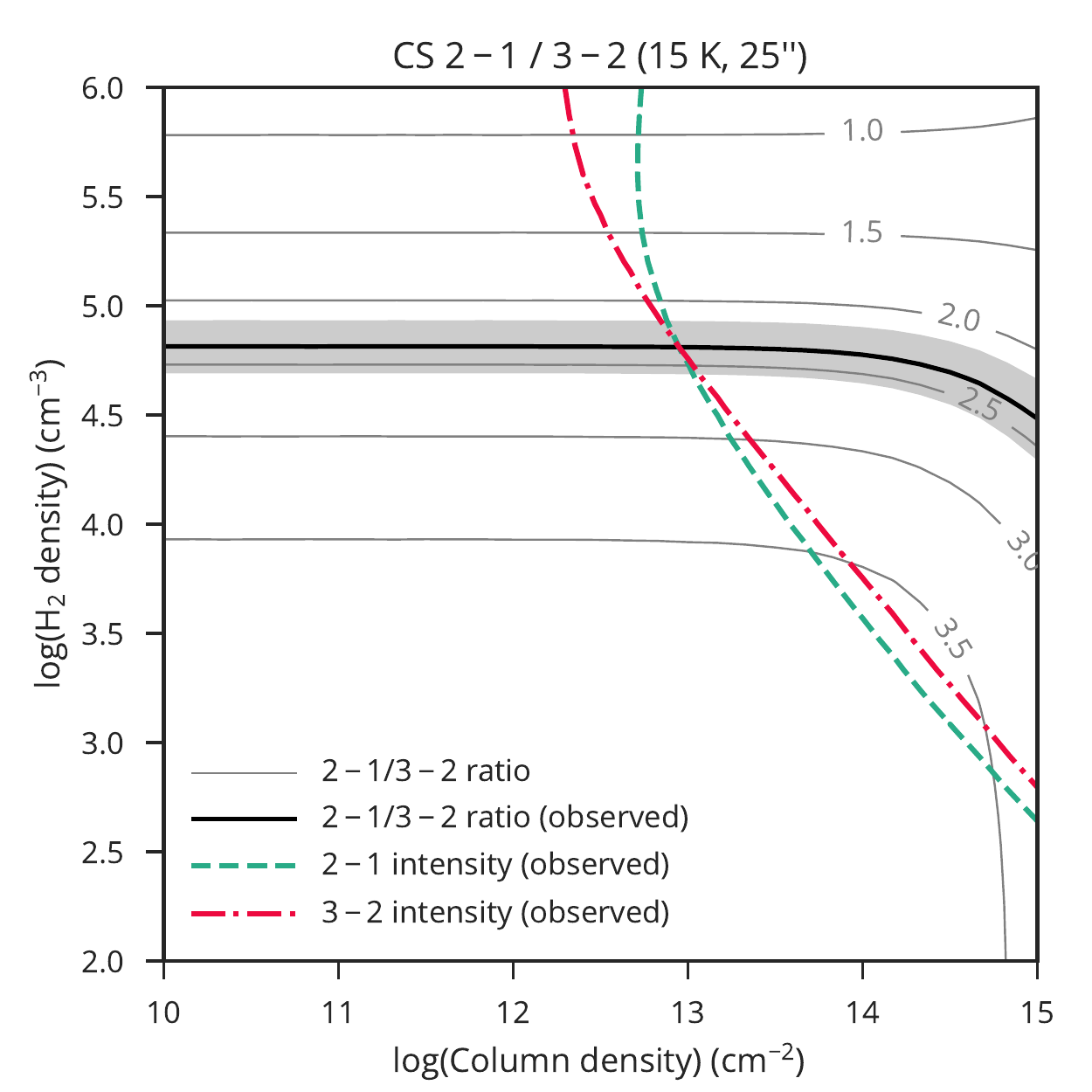}
\caption{RADEX non-LTE model assuming a single density component. 
Model grids are the column density and the H$_2$ density. 
(\emph{left}) The H$_2$CO $1_{01}-0_{00}$/$2_{02}-1_{01}$ 
integrated intensity ratio (black solid), 
the $1_{01}-0_{00}$ integrated intensity (green dashed), 
and the $2_{02}-1_{01}$ integrated intensity (red dash-dotted). 
(\emph{right}) The CS $2-1$/$3-2$ integrated intensity ratio (black solid), 
the $2-1$ integrated intensity (green dashed), 
and the $3-2$ integrated intensity (red dash-dotted). 
The uncertainties of the line ratio (1$\sigma$) are shaded in gray. 
A case for the gas kinetic temperature of 15 K is shown. }
\label{fig:RADEX}
\end{figure*}

As a result, we successfully constrain the H$_2$ densities of the 
emitting regions and the column densities from the observed integrated 
intensities and their ratios for some combinations of the assumed parameters. 
The derived H$_2$ densities and column densities are listed in Table 
\ref{tab:RADEX}. In Table \ref{tab:RADEX}, we also report the optical 
depths for the H$_2$CO ($1_{01}-0_{00}$) and CS ($2-1$) lines for reference. 
Figure \ref{fig:RADEX} shows an example of a plausible 
parameter set. Depending on the gas kinetic temperature ($10-20$ K) 
and the source size ($20-25$\arcsec), the H$_2$ density spans from 
$5.7\times10^3$ to $2.6\times10^4$ cm$^{-3}$ for H$_2$CO 
and $2.9\times10^4$ to $1.2\times10^5$ cm$^{-3}$ for CS. 
The column densities are in the range of $(3.3-9.0)\times10^{12}$ cm$^{-2}$ 
for H$_2$CO\footnote{Note that this is just for para-H$_2$CO. 
If we assume the ortho-to-para ratio to be 3 (the statistical value), 
the total (ortho and para) column density of H$_2$CO is 
in the range of $(1.3-3.6)\times10^{13}$ cm$^{-2}$.} 
and $(0.8-1.7)\times10^{13}$ cm$^{-2}$ for CS. 
Note that the observed intensities for H$_2$CO cannot be reproduced 
with any set of the H$_2$ density and the column density, if the gas 
kinetic temperature is assumed to be of 15 and 20 K with the source 
size of {20\arcsec}. 

\begin{deluxetable*}{@{\extracolsep{8pt}}cccccccc}
\tabletypesize{\scriptsize}
\tablecaption{Results of RADEX non-LTE modeling 
(single density component) \label{tab:RADEX}}
\tablehead{
\colhead{} & \multicolumn2c{Assumed} & \colhead{}
 & \multicolumn3c{Derived} & \colhead{} \\
 \cline{2-3} \cline{5-7}
\colhead{Molecule} & \colhead{$\theta_\mathrm{source}$}
 & \colhead{$T_\mathrm{kin}$}
 & \colhead{} & \colhead{$N_\mathrm{mol}$} 
 & \colhead{$\tau$} 
 & \colhead{$n_\mathrm{H_2}$} & \colhead{Comments} \\
\colhead{} & \colhead{(\arcsec)} & \colhead{(K)} & \colhead{}
 & \colhead{(cm$^{-2}$)} & \colhead{} & \colhead{(cm$^{-3}$)} & \colhead{} \\
\colhead{(1)} & \colhead{(2)} & \colhead{(3)} & \colhead{} 
 & \colhead{(4)} & \colhead{(5)} & \colhead{(6)} & \colhead{(7)}
}
\startdata
H$_2$CO & 25 & 10 & & $3.3\times10^{12}$ & 0.014 & $2.6\times10^{4}$ & \\
        &    & 15 & & $5.0\times10^{12}$ & 0.024 & $1.1\times10^{4}$ 
                                                   & See Figure \ref{fig:RADEX} \\
        &    & 20 & & $7.1\times10^{12}$ & 0.038 & $5.7\times10^{3}$ & \\
        & 20 & 10 & & $9.0\times10^{12}$ & 0.048 & $1.1\times10^{4}$ & \\
        &    & 15 & &         --         &   --   &    --    & \tablenotemark{a} \\
        &    & 20 & &         --         &   --   &    --    & \tablenotemark{a} \\
\hline
CS      & 25 & 10 & & $8.1\times10^{12}$ & 0.014 & $1.2\times10^{5}$ & \\
        &    & 15 & & $8.9\times10^{12}$ & 0.016 & $6.4\times10^{4}$ 
                                                   & See Figure \ref{fig:RADEX} \\
        &    & 20 & & $1.0\times10^{13}$ & 0.019 & $4.3\times10^{4}$ & \\
        & 20 & 10 & & $1.2\times10^{13}$ & 0.022 & $9.5\times10^{4}$ & \\
        &    & 15 & & $1.4\times10^{13}$ & 0.028 & $4.7\times10^{4}$ & \\
        &    & 20 & & $1.7\times10^{13}$ & 0.036 & $2.9\times10^{4}$ & 
\enddata
\tablecomments{Columns are: (1) molecular species, 
(2) assumed source size, (3) assumed gas kinetic temperature, 
(4) derived column density, (5) derived optical depth of the 
$1_{01}-0_{00}$ and $2-1$ lines for H$_2$CO and CS, respectively, 
(6) derived H$_2$ density of the emitting region, (7) other remarks.}
\tablenotetext{a}{The assumed parameter set of the gas kinetic temperature 
and the source size does not reproduce the observed intensities 
at any set of the H$_2$ density and the column density. }
\end{deluxetable*}

\subsection{The H$_2$ densities for the emitting 
regions of H$_2$CO and CS} \label{subsec:effective}

As described in the previous section, the H$_2$ densities for the 
emitting region of H$_2$CO and CS are evaluated to be $(0.6-2.6)\times10^4$ 
cm$^{-3}$ and $(2.9-12)\times10^4$ cm$^{-3}$, respectively. 
It seems reasonable that the derived H$_2$ densities for the emitting 
region of H$_2$CO ($1_{01}-0_{00}$) and CS ($2-1$) 
are higher than that derived for $^{13}$CO ($1-0$) and $^{12}$CO 
($1-0$, $2-1$) \citep[$120-240$ cm$^{-3}$;][]{Schinnerer2010}, 
because the H$_2$CO and CS lines have higher critical densities than 
the CO isotoplogue lines. 
Indeed, the derived densities are similar to the effective critical 
densities \citep[H$_2$CO ($1_{01}-0_{00}$): $3\times10^4$ cm$^{-3}$, 
CS ($2-1$): $5\times10^4$ cm$^{-3}$ at 15 K;][]{Shirley2015}%
\footnote{The effective critical density is the density at which 
collisional and radiative processes balance each other under taking 
radiative trapping into account. It depends on the molecular column density. 
The above mentioned values are scaled to the derived 
column densities by using Eq.~(13) of \citet{Shirley2015}.}. 

As expected, the derived density in the spiral arm of M51 is lower than 
that in the central region of M51 derived by the HCN ($1-0$) and $^{13}$CO 
($1-0$) lines \citep[$>100$ K and $\sim10^5$ cm$^{-3}$;][]{Matsushita1998}, 
where a heavily-obscured AGN associated with radio jet and ionized/molecular 
outflows exists \citep[e.g.,][]{Kohno1996, Matsushita2007}. 
This result seems reasonable, because the observed position in the 
spiral arm is, unlike the central region, free from nuclear jets 
and/or compression by supernova-driven winds. 

Although the H$_2$ densities of the emitting region 
of H$_2$CO and CS are higher than that of CO isotopologues, they seem to be 
widely distributed over the GMCs. If H$_2$CO and CS are localized 
to dense clumps and are deficient in diffuse parts of clouds, the derived gas 
densities would be higher. The similarity of the derived H$_2$ densities 
and the effective critical densities may imply that H$_2$CO and CS reside in 
the wide ranges of density from that well below to that well above 
the effective critical densities. Hence, both H$_2$CO and CS are not 
localized to dense star-forming cores ($<0.1$ pc and $>10^5$ cm$^{-3}$), 
but rather reside in a considerably large fraction of the GMCs. Our result 
indicates that H$_2$CO and CS in relatively less dense parts of molecular 
clouds mainly contribute to their emission in the 3 mm and 2 mm regions. 
This result should be considered in interpretation of molecular emissions 
in a disk region of a galaxy observed with a molecular-cloud-scale beam. 

\section{Implications for realistic density distributions}
\label{sec:distribution}

Both theoretical and observational studies have pointed out that 
molecular clouds comprise a wide range of gas densities from diffuse to dense ones. 
As a functional form of density distribution, lognormal functions are often 
predicted by turbulent theories \citep[e.g.,][]{Padoan2002}. Milky Way 
observations are basically consistent with these theories, but it is sometimes 
pointed out that power laws better describe distributions at high densities 
\citep[e.g.,][]{Lombardi2015}. To explore the density distribution from 
spatially-unresolved observations toward external galaxies, \citet{Leroy2017} 
demonstrated that a proper set of molecular line intensities is useful to 
probe underlying distributions. Given that H$_2$CO and CS are widely 
distributed over the molecular clouds, we here consider the H$_2$ density 
distribution within the observed 1 kpc beam, assuming a lognormal distribution 
with and without a power law tail as employed by \citet{Leroy2017}.

\subsection{Modeling of line intensity and emissivity}
\label{subsec:LeroyRADEX}

\begin{figure}
\centering
\includegraphics[width=0.9\hsize]{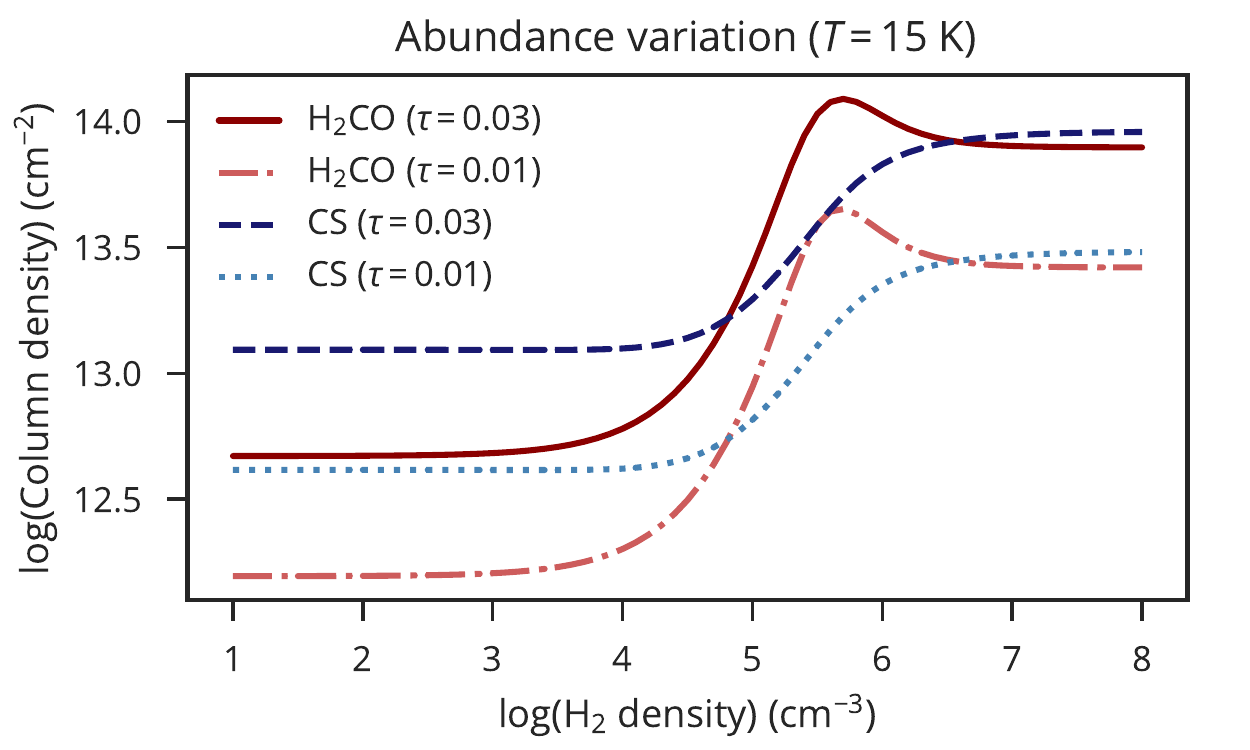}
\caption{Variation of the column densities of H$_2$CO and CS 
within the model H$_2$ density distribution under the fixed optical depth 
for the lower-$J$ transition. Here, we fix the gas kinetic temperature to be 15 K. 
Cases for the optical depths of $\tau=0.03$ and $0.01$ are shown. 
Note that these results do not take any chemical processes into account.}
\label{fig:abundance}
\end{figure}

Prior to integrating the intensities for a wide range of H$_2$ density, 
we run the RADEX code to calculate line intensities of the H$_2$CO and CS 
transitions at each density. As the input parameters for RADEX, we adopt 
the same values as Section \ref{sec:analyses}: the background temperature of 
2.73 K, the gas kinetic temperature $T_\mathrm{kin}$ of 10 and 15 K%
\footnote{Here, we do not consider the case of $T_\mathrm{kin}=$ 20 K. 
At 20 K, the level populations of the $0_{00}$ and $1_{01}$ levels 
of H$_2$CO are inverted in a certain range of the H$_2$ density and 
a certain range of the column density. To keep the optical depth fixed, 
the column density of H$_2$CO has to be discontinuous as a function of H$_2$ density. 
Such a discontinuity of column density is not realistic.}, 
and line widths of 37 and 48 km s$^{-1}$ for H$_2$CO and CS, respectively. 
The H$_2$ densities are 70 values logarithmically spaced in the range of 
from $10^1$ to $10^8$ cm$^{-3}$. Following \citet{Leroy2017}, 
we adjust the molecular column density to keep the optical depth fixed. 
Based on the estimation in Section \ref{subsec:RADEX}, we set 
the optical depth of the lower-$J$ transitions to be $\tau=$ 0.03 and 0.01. 
The optical depth of higher-$J$ transition is calculated consistently 
from the lower-$J$ transition. Under the fixed optical depth, 
the column density varies as a function of H$_2$ density (Figure \ref{fig:abundance}). 
However, this abundance variations of H$_2$CO and CS may be underestimated, 
because chemical models predict even larger variations along H$_2$ density 
\citep[e.g.,][]{Harada2019}. The abundance variations could be 
improved by taking chemical processes into account in future work.

For each density step, we calculate emissivity $\epsilon$. Here, 
emissivity is defined as the intensity $I$ divided by the H$_2$ column density 
$N_\mathrm{H_2}$. For consistency with \citet{Leroy2017}, the column density of 
H$_2$ is calculated by dividing the column density of emitting molecule 
$N_\mathrm{mol}$, by the fractional abundance of the molecule $X_\mathrm{mol}$. 
Then, the emissivity is calculated as $\epsilon = I/N_\mathrm{mol}X_\mathrm{mol}^{-1}$. 
The fractional abundance is adjusted to be $10^{-9}$ for both H$_2$CO and CS 
so that $N_\mathrm{mol}X_\mathrm{mol}^{-1}$ is roughly comparable with 
the reported column density of $N_\mathrm{H_2} = (2.3-3.3)\times10^{22}$ cm$^{-2}$ 
\citep{Schinnerer2013}.

\subsection{Lognormal and power law distributions}
\label{subsec:lognormal}

Following \citet{Leroy2017}, we consider two kinds of functional form 
for the H$_2$ density distribution: lognormals with and without a power law tail 
at high density. Lognormal distributions are described by: 
$$
dP(\ln n')\propto\exp\left(-\frac{(\ln n'-\overline{\ln n'})^2}{2\sigma^2}\right)d\ln n',
$$
where $dP$ is the fraction of cells in a logarithmic step $d\ln n'$, 
$n' = n_\mathrm{H_2}/n_0$ is the H$_2$ density normalized by the mean H$_2$ 
density $n_0$, and $\sigma$ is the rms dispersion of the distribution. 
The distribution peaks at the mean of the logarithm of the density $\overline{\ln n'}$, 
which is determined by the relation: $\overline{\ln n'} = -\sigma^2/2$. 
As the lognormal distribution is a result from isothermal supersonic turbulence, 
$\sigma$ is related to the turbulent Mach number $\mathcal{M}$: 
$\sigma^2 = \ln(1+\mathcal{M}^2/4)$ \citep[for more details, see][]{Padoan2002}. 
For easier comparison with \citet{Leroy2017}, we use $\log_{10}$, so that $\sigma$ 
is represented in dex. To incorporate a power law tail, we employ the formulation: 
$$
dP(\ln n') \propto \exp\left(\alpha\ln n'\right)\ \textrm{for}\ n'>n'_\mathrm{thresh},
$$
where the power law index is taken to be $\alpha=-1.5$, and the threshold for 
the power law tail to be $\ln n'_\mathrm{thresh}=3.8$ \citep{Federrath2013}. 
The model grids of the mean H$_2$ density and the width of distribution are set 
to be logarithmically spaced 40 points in the range of $n_0=10^2-10^6$ cm$^{-3}$, 
and linearly spaced 8 points in the range of $\sigma=0.4-1.2$, respectively, 
for each of lognormal distributions with and without power law tails. 

\subsection{Realistic H$_2$ density distributions}
\label{subsec:realistic}

For each H$_2$ density distribution, we derive the beam averaged emissivity 
by summing up the emissivity along the axis of H$_2$ density \citep{Leroy2017}:
$$
\langle\epsilon\rangle
=\frac{\int n_\mathrm{H_2}\,P(n_\mathrm{H_2})\,%
\epsilon(n_\mathrm{H_2},T_\mathrm{kin},\tau)\,dn_\mathrm{H_2}}%
{\int n_\mathrm{H_2}\,P(n_\mathrm{H_2})\,dn_\mathrm{H_2}}.
$$
We calculate the line intensities integrated over the 1 kpc beam, 
by multiplying the H$_2$ column density of $N_\mathrm{H_2}=3\times10^{22}$ 
cm$^{-2}$ \citep{Schinnerer2013} to the beam averaged emissivity 
$\langle\epsilon\rangle$. This causes minor inconsistency between 
$N_\mathrm{H_2}$ and $N_\mathrm{mol}X_\mathrm{mol}^{-1}$, because 
$N_\mathrm{mol}X_\mathrm{mol}^{-1}$ slightly vary within the model distribution.
As we do not know the exact fractional abundances of molecules at a specific 
H$_2$ density, we employ this simplest assumption.

\begin{figure*}
\centering
\includegraphics[width=0.45\hsize]{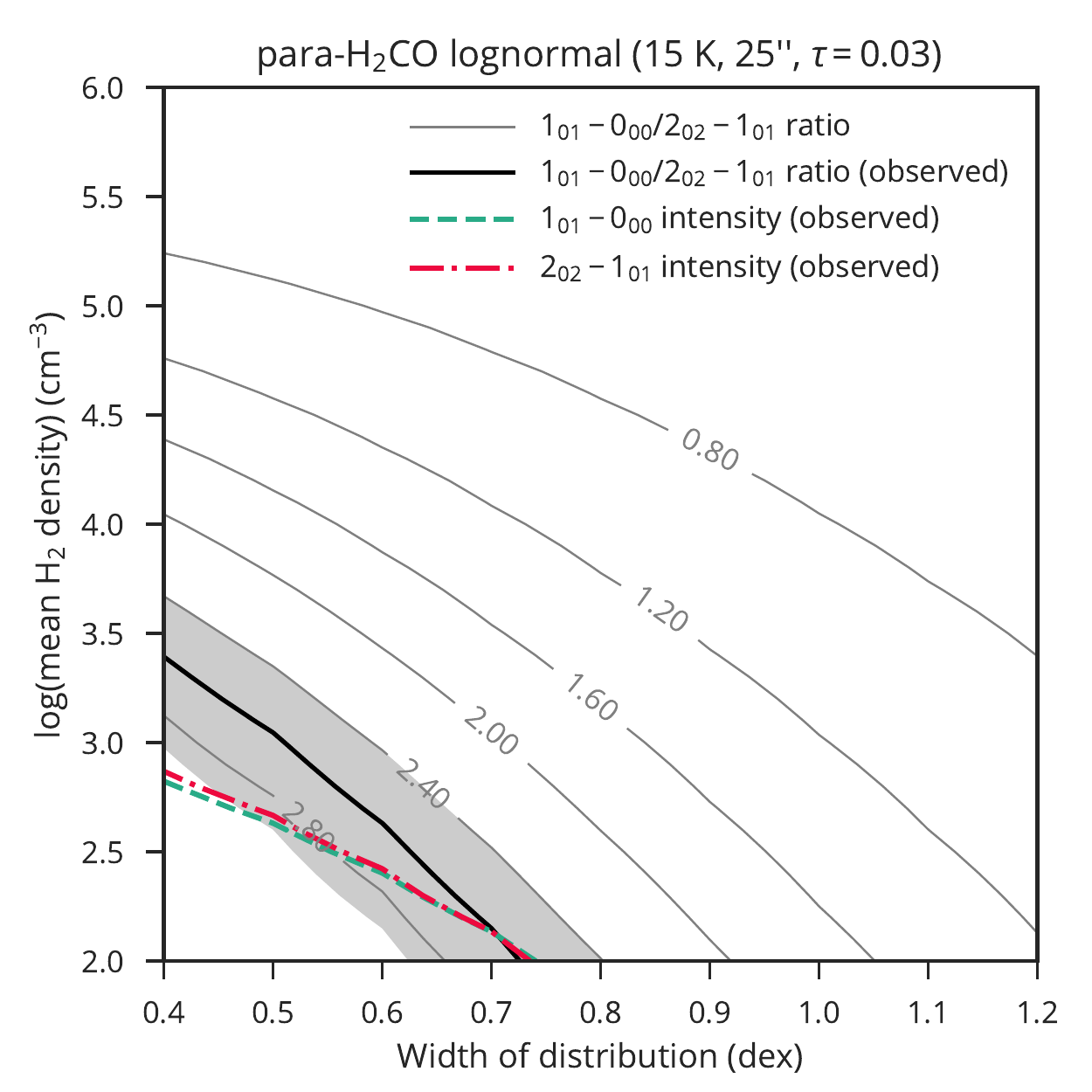}\hspace{0.03\hsize}
\includegraphics[width=0.45\hsize]{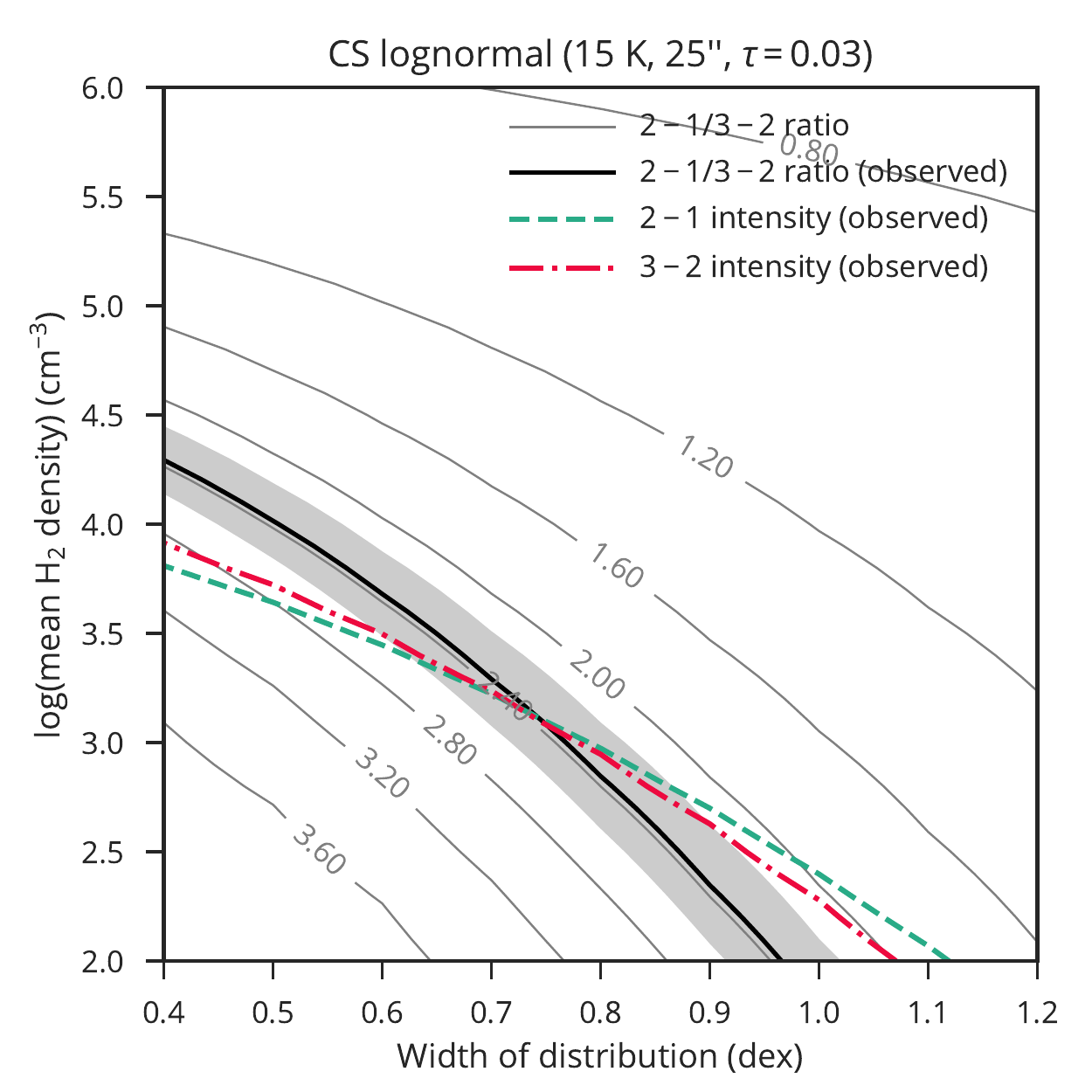}\\
\includegraphics[width=0.45\hsize]{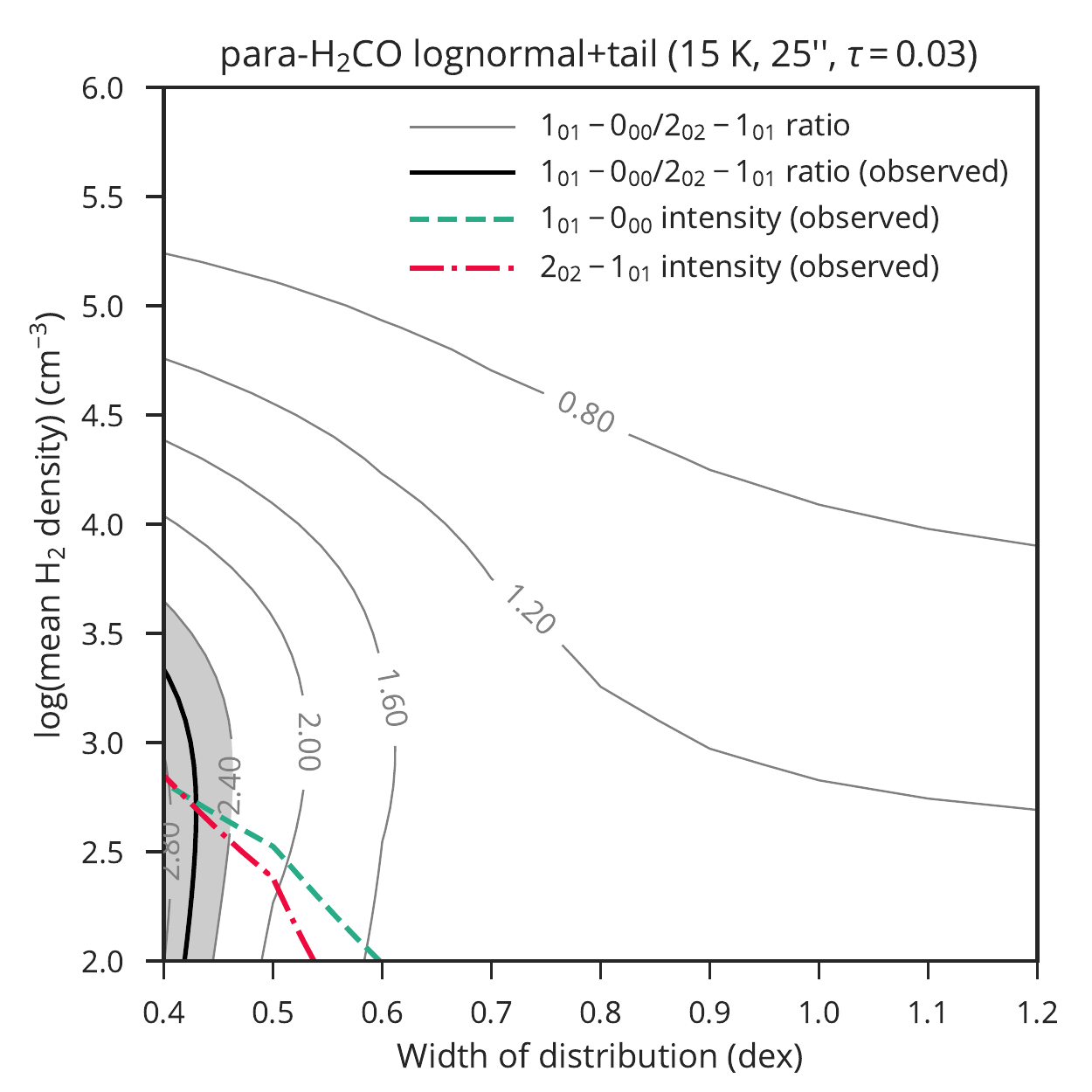}\hspace{0.03\hsize}
\includegraphics[width=0.45\hsize]{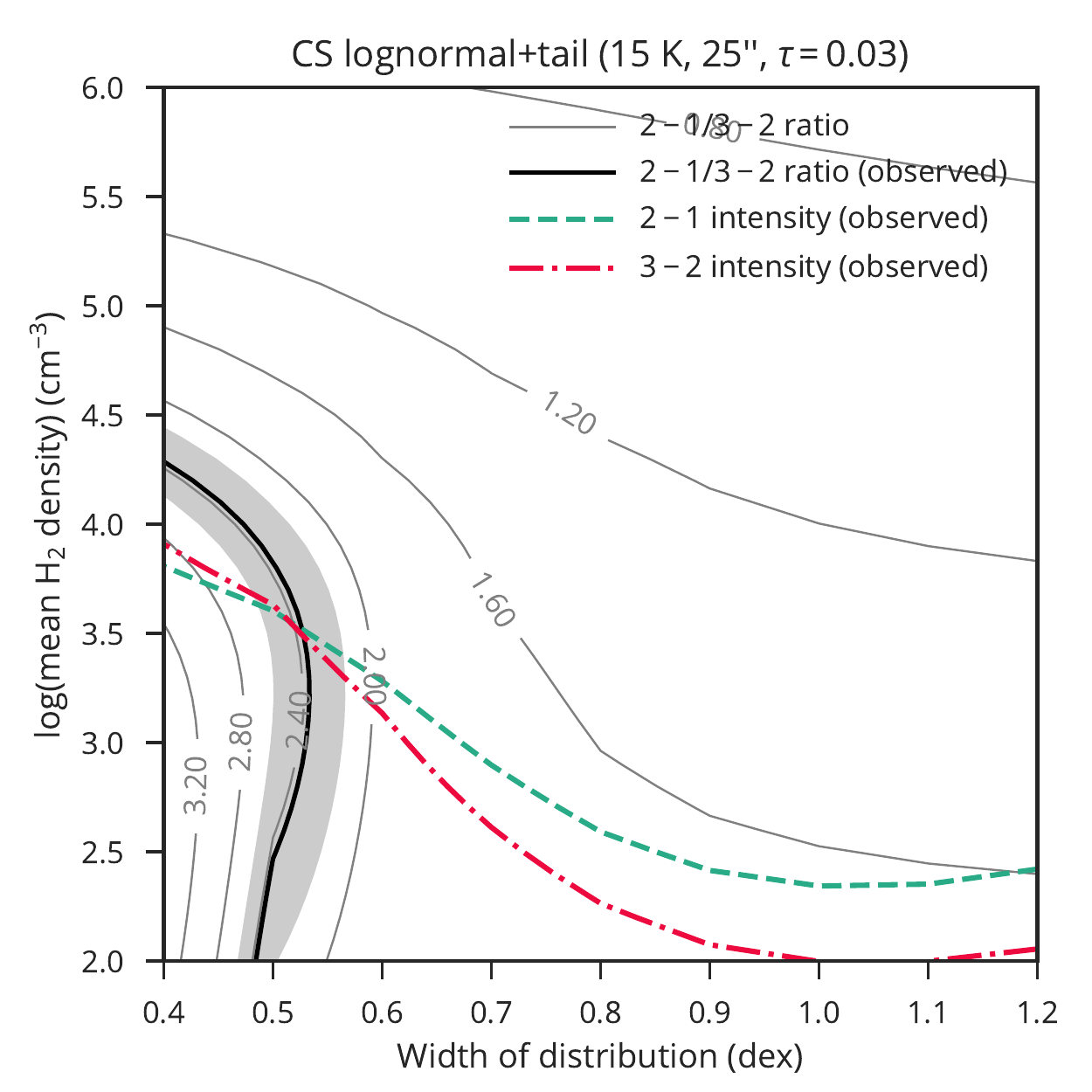}
\caption{As in Figure \ref{fig:RADEX}, but assuming the lognormal 
distribution of the H$_2$ densities with (\emph{upper panels}) and without 
(\emph{lower panels}) a power law tail at high density. 
Model grids are the mean H$_2$ density and the rms dispersion of the lognormal 
distribution. Here, we adopt the fixed optical depth of lower-$J$ transitions 
($\tau=0.03$). In the cases with the power law tail, the intensity ratio patterns 
differ from the pure lognormal distributions at large widths of the distribution. }
\label{fig:lognormal}
\end{figure*}

\begin{figure*}
\centering
\includegraphics[width=0.45\hsize]{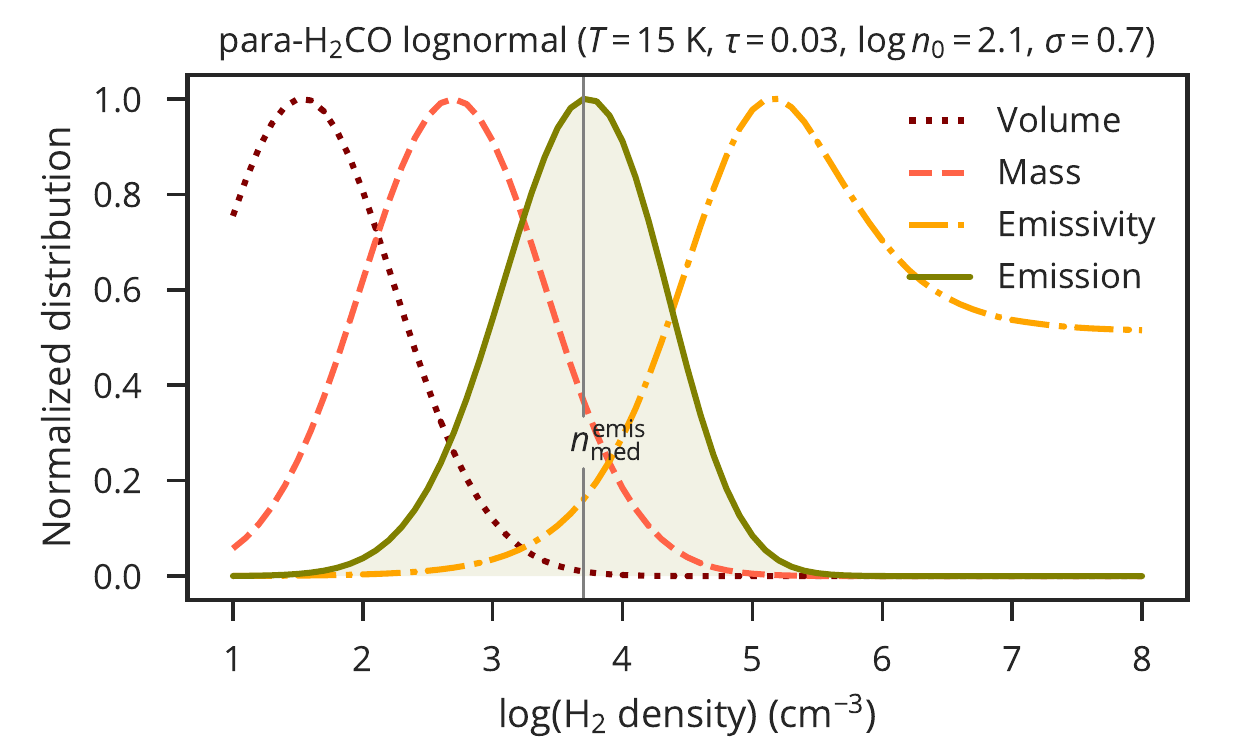}\hspace{0.03\hsize}
\includegraphics[width=0.45\hsize]{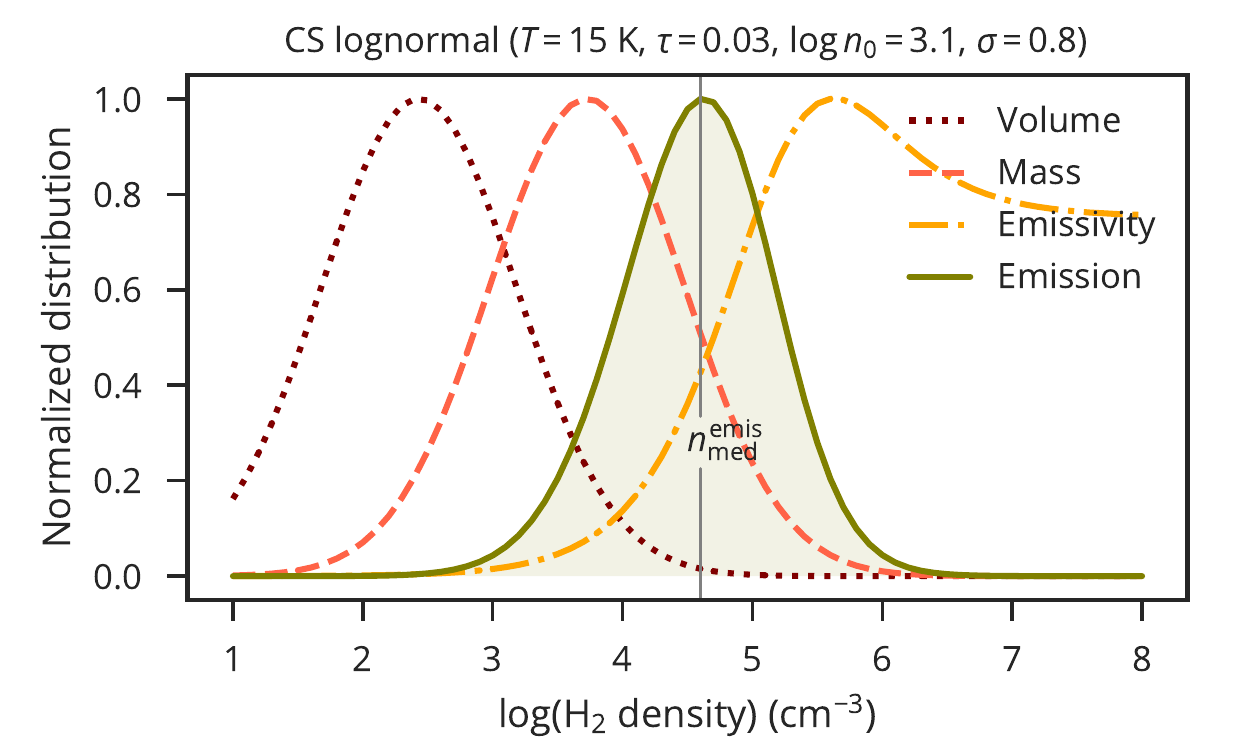}\\
\includegraphics[width=0.45\hsize]{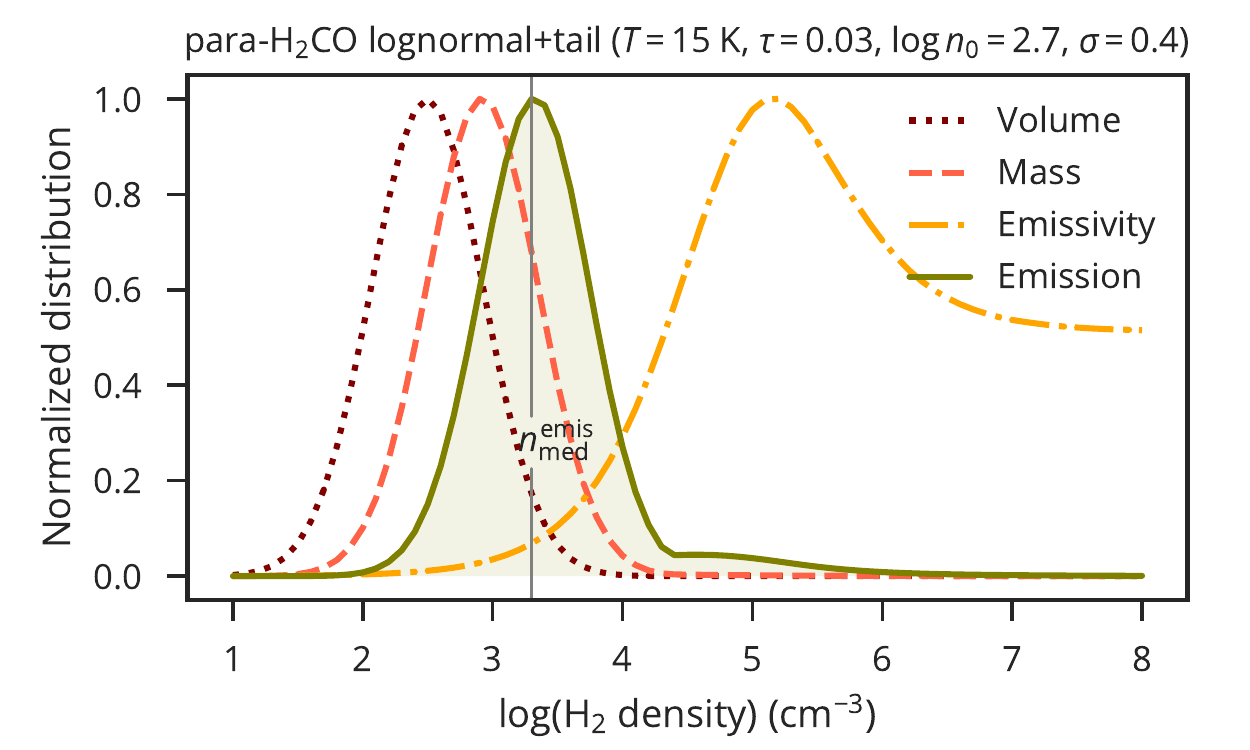}\hspace{0.03\hsize}
\includegraphics[width=0.45\hsize]{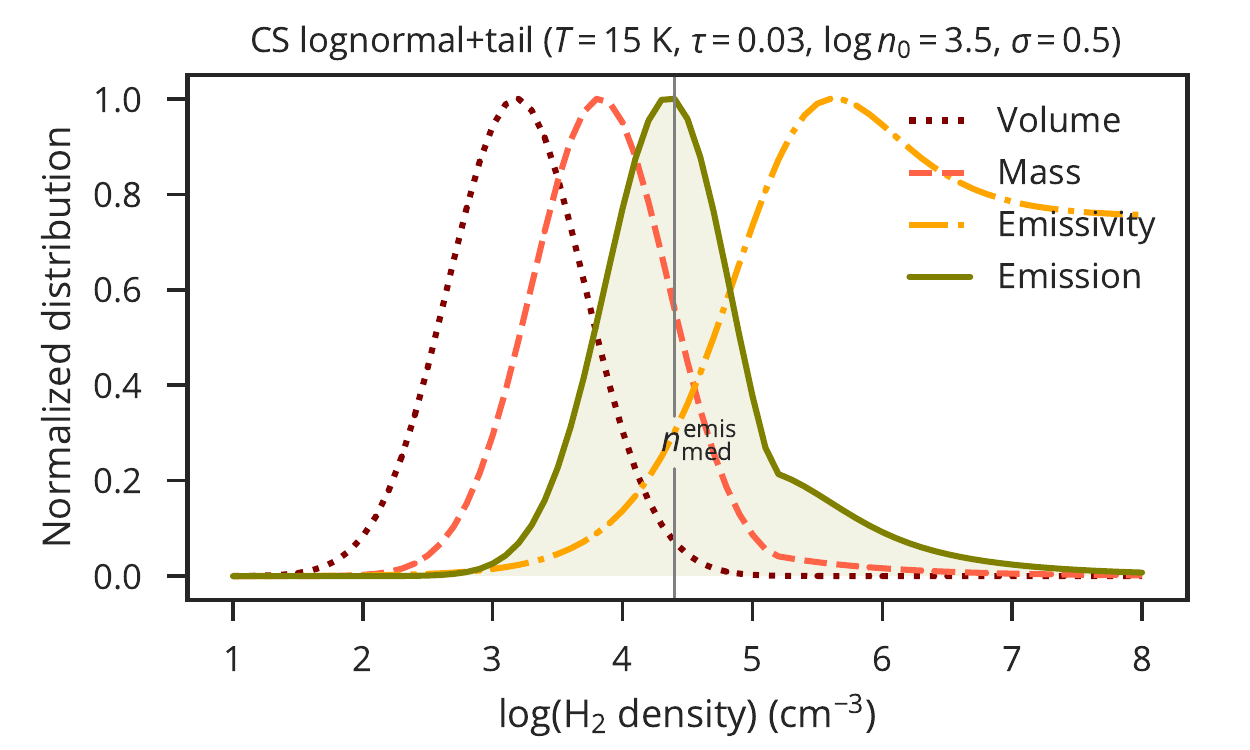}
\caption{Model distributions of H$_2$ density by volume (dotted) and 
mass (dashed), molecular emissivity (dash-dotted), and emission (solid) 
for plausible parameter sets in Figure \ref{fig:lognormal}. 
Each distribution is normalized to its peak value at any H$_2$ density. 
Vertical lines indicate the median H$_2$ density for molecular emission. 
Here, molecular emissivity is defined as emission per unit column density. }
\label{fig:distribution}
\end{figure*}

By comparing the modeled line intensities and their ratios with observed 
ones, the parameters of the density distribution are constrained as summarized in 
Table \ref{tab:lognormal}. In Figure \ref{fig:lognormal}, we show the model grids 
for plausible parameter sets. Figure \ref{fig:distribution} illustrates 
the distribution of volume, mass, emissivity and emission for each plausible 
parameter sets. Compared with pure lognormal distributions, the model with 
a power law tail tends to show the lower mean H$_2$ density and the narrower 
width of distribution. The mean H$_2$ density 
is in a range of $n_0=130-6600$ cm$^{-3}$. This seems reasonable, because 
the range overlaps to $120-240$ cm$^{-3}$, which is derived from $^{12}$CO 
and $^{13}$CO observations \citep{Schinnerer2013}. The width of distribution, 
on the other hand, is in the range of $\sigma=0.4-0.9$, which corresponds to 
$\mathcal{M}=2.3-18$ in turbulent Mach number. This range almost falls on the 
smallest case of the typical value ($5-100$) for spiral and starburst galaxies 
\citep{Leroy2016}.

Note that, in any density distribution, we cannot simultaneously 
reproduce the observed intensity of H$_2$CO and CS. This would probably 
originate from the fixed kinetic temperature and, most notably, the 
arbitrary assumption for their abundances to keep the optical depth fixed. 
In the next section, we discuss chemical processes to form H$_2$CO and CS, 
which may differentiate their distribution. Implementing more realistic 
molecular abundance variation is awaited for future study. 

Finally, we discuss the median intensity for emission 
$n_\mathrm{med}^\mathrm{emis}$, which is defined by the H$_2$ density 
below which 50\% of the line emission emerges \citep{Leroy2017}. 
Here, we consider the lines of lower-$J$ transitions, i.e., $1_{01}-0_{00}$ 
for H$_2$CO and $2-1$ for CS. The modeled values of the median intensity 
for emission are shown in Table \ref{tab:lognormal}. These are consistent 
with the H$_2$ densities of the emitting regions derived 
in Section \ref{sec:analyses}. This supports the picture suggested in 
Section \ref{subsec:effective} that H$_2$CO and CS are present 
in a wide range of H$_2$ density.

\begin{deluxetable*}{@{\extracolsep{8pt}}cccccccccc}
\tabletypesize{\scriptsize}
\tablecaption{Results of RADEX non-LTE modeling 
(lognormal distribution with and without power law tail) \label{tab:lognormal}}
\tablehead{
\colhead{} & \multicolumn4c{Assumed} & \colhead{}
 & \multicolumn3c{Derived} & \colhead{} \\
 \cline{2-5} \cline{7-9}
\colhead{Molecule} & \colhead{$\theta_\mathrm{source}$}
 & \colhead{$T_\mathrm{kin}$}  & \colhead{$\tau$}  & \colhead{Distribution} 
 & \colhead{} & \colhead{$n_0$}  & \colhead{$\sigma$} 
 & \colhead{$n_\mathrm{med}^\mathrm{emis}$} & \colhead{Comments} \\
\colhead{} & \colhead{(\arcsec)} & \colhead{(K)} & \colhead{} & \colhead{}
 & \colhead{} & \colhead{(cm$^{-3}$)} & \colhead{(dex)} & \colhead{(cm$^{-3}$)} 
 & \colhead{} \\
\colhead{(1)} & \colhead{(2)} & \colhead{(3)} & \colhead{(4)} & \colhead{(5)} 
 & \colhead{} & \colhead{(6)} & \colhead{(7)} & \colhead{(8)} & \colhead{(9)}
}
\startdata
H$_2$CO & 25 & 10 & 0.01 & lognormal & & -- & -- & -- & \tablenotemark{a} \\
        &    &    & 0.03 & lognormal & & -- & -- & -- & \tablenotemark{a} \\
        &    & 15 & 0.01 & lognormal & & $1.3\times10^{2}$ & 0.7 & $5.0\times10^{3}$ & \\
        &    &    & 0.03 & lognormal & & $1.3\times10^{2}$ & 0.7 & $5.0\times10^{3}$ 
                           & See Figure \ref{fig:lognormal} and \ref{fig:distribution} \\
        & 20 & 10 & 0.01 & lognormal & & $6.5\times10^{2}$ & 0.5 & $6.3\times10^{3}$ & \\
        &    &    & 0.03 & lognormal & & $6.5\times10^{2}$ & 0.5 & $6.3\times10^{3}$ & \\
        &    & 15 & 0.01 & lognormal & & -- & -- & -- & \tablenotemark{a} \\
        &    &    & 0.03 & lognormal & & -- & -- & -- & \tablenotemark{a} \\
\hline
CS      & 25 & 10 & 0.01 & lognormal & & $6.8\times10^{2}$ & 0.9 & $7.9\times10^{4}$ & \\
        &    &    & 0.03 & lognormal & & $7.4\times10^{2}$ & 0.9 & $7.9\times10^{4}$ & \\
        &    & 15 & 0.01 & lognormal & & $1.2\times10^{3}$ & 0.7 & $4.0\times10^{4}$ & \\
        &    &    & 0.03 & lognormal & & $1.2\times10^{3}$ & 0.7 & $4.0\times10^{4}$ 
                           & See Figure \ref{fig:lognormal} and \ref{fig:distribution} \\
        & 20 & 10 & 0.01 & lognormal & & $4.0\times10^{3}$ & 0.7 & $6.3\times10^{4}$ & \\
        &    &    & 0.03 & lognormal & & $4.1\times10^{3}$ & 0.7 & $6.3\times10^{4}$ & \\
        &    & 15 & 0.01 & lognormal & & $5.2\times10^{3}$ & 0.5 & $3.2\times10^{4}$ & \\
        &    &    & 0.03 & lognormal & & $5.3\times10^{3}$ & 0.5 & $3.2\times10^{4}$ & \\
\hline
H$_2$CO & 25 & 10 & 0.01 & lognormal+tail & & $5.6\times10^{2}$ & 0.5 & $3.2\times10^{3}$ & \\
        &    &    & 0.03 & lognormal+tail & & $5.7\times10^{2}$ & 0.5 & $3.2\times10^{3}$ & \\
        &    & 15 & 0.01 & lognormal+tail & & $5.1\times10^{2}$ & 0.4 & $2.0\times10^{3}$ & \\
        &    &    & 0.03 & lognormal+tail & & $5.1\times10^{2}$ & 0.4 & $2.0\times10^{3}$ 
                                & See Figure \ref{fig:lognormal} and \ref{fig:distribution} \\
        & 20 & 10 & 0.01 & lognormal+tail & & $1.2\times10^{3}$ & 0.4 & $4.0\times10^{3}$ & \\
        &    &    & 0.03 & lognormal+tail & & $1.2\times10^{3}$ & 0.4 & $4.0\times10^{3}$ & \\
        &    & 15 & 0.01 & lognormal+tail & & -- & -- & -- & \tablenotemark{a} \\
        &    &    & 0.03 & lognormal+tail & & -- & -- & -- & \tablenotemark{a} \\
\hline
CS      & 25 & 10 & 0.01 & lognormal+tail & & $3.1\times10^{3}$ & 0.6 & $4.0\times10^{4}$ & \\
        &    &    & 0.03 & lognormal+tail & & $3.2\times10^{3}$ & 0.6 & $4.0\times10^{4}$ & \\
        &    &    & 0.01 & lognormal+tail & & $3.2\times10^{3}$ & 0.5 & $2.5\times10^{4}$ & \\
        &    &    & 0.03 & lognormal+tail & & $3.2\times10^{3}$ & 0.5 & $2.5\times10^{4}$ 
                                & See Figure \ref{fig:lognormal} and \ref{fig:distribution} \\
        & 20 & 10 & 0.01 & lognormal+tail & & $6.6\times10^{3}$ & 0.5 & $5.0\times10^{4}$ & \\
        &    &    & 0.03 & lognormal+tail & & $6.7\times10^{3}$ & 0.5 & $5.0\times10^{4}$ & \\
        &    & 15 & 0.01 & lognormal+tail & & $6.5\times10^{3}$ & 0.5 & $2.5\times10^{4}$ & \\
        &    &    & 0.03 & lognormal+tail & & $6.6\times10^{3}$ & 0.5 & $2.5\times10^{4}$ & \\
\enddata
\tablecomments{Columns are: (1) molecular species, 
(2) assumed source size, (3) assumed gas kinetic temperature, 
(4) assumed optical depth of the $1_{01}-0_{00}$ and $2-1$ lines 
for H$_2$CO and CS, respectively, (5) assumed density distribution, 
(6) mean H$_2$ density for the lognormal part of the density distribution, 
(7) dispersion in the lognormal part of the density distribution, 
(8) median density below which 50\% of the emission of the lower-$J$ transition emerges, 
(9) other remarks.}
\tablenotetext{a}{The assumed parameter set of the gas kinetic temperature, 
the source size, and the optical depth does not reproduce the observed intensities 
at any set of the mean H$_2$ density and the width of distribution.}
\end{deluxetable*}

\section{Implications for chemical processes} \label{sec:chemistry}

Formation and destruction processes of molecules depend on the physical 
conditions. The relatively low density ($\sim10^4$ cm$^{-3}$) of the 
emitting regions of H$_2$CO and CS suggests that these molecules are 
not localized to dense and hot regions associated with active star 
formation. In the following subsections, implications of this result to 
the chemical processes of H$_2$CO and CS are discussed. 

\subsection{H$_2$CO} \label{subsec:H2CO}

In general, both grain surface reactions and gas phase reactions 
are thought to contribute to the production of H$_2$CO. 
Theoretical models and laboratory experiments suggest that 
successive hydrogenation of CO in icy mantles of dust grains 
followed by subsequent liberation into the gas phase by thermal 
and/or non-thermal desorption is the dominant process 
\citep[CO $\rightarrow$ HCO $\rightarrow$ H$_2$CO;][]{Watanabe2002}, 
although the gas phase formation can produce some abundance of 
H$_2$CO \citep[e.g.,][]{Soma2018}. 

In our case, widespread H$_2$CO in relatively less dense gas would not 
originate from thermal desorption. As the desorption temperature is 
as high as 40 K, the thermal desorption is restricted in small 
hot and dense regions around newly born stars (i.e., hot cores) 
and/or outflow shocked regions. Then, the non-thermal desorption should 
mainly be responsible for the liberation of H$_2$CO formed in ice mantles. This is supported by the previous observation of CH$_3$OH toward 
this position of M51 \citep{Watanabe2014, Watanabe2016}. According to 
them, the distribution of CH$_3$OH is widespread over a 100 pc scale. 
CH$_3$OH is formed on grain surface by further hydrogenation of H$_2$CO 
\citep[H$_2$CO $\rightarrow$ CH$_3$O $\rightarrow$ CH$_3$OH;][]%
{Watanabe2002}, whereas it is not efficiently produced by the 
gas-phase reactions \citep{Geppert2006}. Therefore, a widespread 
CH$_3$OH suggests its non-thermal desorption, and in this case, 
H$_2$CO should also be liberated. As for the non-thermal desorption 
processes, sputtering of molecules in large-scale shocks such as 
spiral shocks and cloud-cloud collisions as well as temporal heating 
of dust grains by cosmic rays and liberation assisted by surplus 
reaction energy in formation of molecules are proposed 
\citep[e.g.,][]{Hasegawa1993,Garrod2007}. It is worth noting that desorption 
by UV photons can also work for H$_2$CO but not for CH$_3$OH 
\citep{Martin-Domenech2016}. 

The abundance ratio CH$_3$OH/H$_2$CO in M51 P1 is evaluated to be 1.5 
by using the column density of H$_2$CO obtained in this study and that 
of CH$_3$OH reported by \citet{Watanabe2014}. Here, the ortho-to-para 
ratio of H$_2$CO is assumed to be the statistical value of 3, and 
the line intensity of CH$_3$OH is corrected for the source size of 
{25\arcsec} for the fair comparison. The CH$_3$OH/H$_2$CO ratio in 
various Galactic objects varies by two orders of magnitude 
from $0.3$ \citep[e.g., $\rho$ Oph A D-peak;][]{Bergman2011} 
to $24$ \citep[e.g., NGC 6334 IRS1;][]{vanderTak2000}. The difference 
may originate from different hydrogenation of CO in ice mantles and/or 
different contribution of the gas phase production. The ratio in M51 P1 
is just in the middle of this range. It is not very different from those 
in the other external galaxies M82 \citep[1.1;][]{Aladro2011b} and 
NGC 253 \citep[5.2;][]{Martin2006}, although the physical condition 
of M51 P1 is much different form these two sources. 

\subsection{CS} \label{subsec:CS}

It is well known that CS is ubiquitous in various interstellar sources 
including diffuse clouds \citep[e.g.,][]{Lucas2002}. Gas-phase reactions 
are thought to be important in its production, and hence, CS can be 
formed under the less dense condition of M51 P1. According to the 
interferometric observation by \citet{Watanabe2016}, the distribution 
of CS is slightly more extended than CH$_3$OH in M51 P1. Hence, the 
different spatial distribution would originate from the contribution 
of the gas-phase production of CS as oppose to the grain-surface 
production of CH$_3$OH. The slightly different H$_2$ densities derived 
from H$_2$CO and CS would suggest the different production mechanism. 

The CS/H$_2$CO ratio is evaluated to be 0.5 in M51 P1. The ratios in 
various Galactic sources range from 0.4 \citep[e.g., Galactic diffuse 
clouds;][]{Liszt2006} to 1.6 \citep[e.g., IRAS 16293-2422;][]{Blake1994,%
vanDishoeck1995}, and hence, the ratio in M51 P1 falls in this range. 
It is worth noting that the ratios are 1.8 in M82 \citep{Aladro2011b} 
and 1.6 (180 km s$^{-1}$ component) -- 2.9 (285 km s$^{-1}$ component) 
in NGC 253 \citep{Martin2006}, which also agree with the ratio in M51 P1 
by a factor of a few, as in the case for the CH$_3$OH/H$_2$CO ratio. 
We need more samples to examine whether this is just by chance or 
by some chemical reasons. 

These results provide us with an additional support that 
the molecular composition averaged over a few $10-100$ pc scale observed 
with single dish telescopes in the 3 mm wavelength range mostly 
reflects that of the widespread gas rather than that affected by 
local star formation activities, as inferred by the previous studies 
of Galactic GMCs \citep{Nishimura2017,Watanabe2017}. According to them, 
the 3 mm-band spectra averaged over a GMC scale (the 39 pc $\times$ 
47 pc area in W51; \citealp{Watanabe2017} and the 9 pc $\times$ 9 pc 
area in W3(OH); \citealp{Nishimura2017}) are similar to the spectra of 
diffuse cloud peripheries rather than those of dense regions. 
This trend is also supported by the chemical model calculation by 
\citet{Harada2019}.

\section{Summary}

We study the gas densities of the emitting regions of the two 
commonly-observed molecular species H$_2$CO and CS in a 1 kpc region 
in the spiral arm of M51. We have conducted sensitive observations of 
the H$_2$CO ($1_{01}-0_{00}$) line using the NRO 45 m and IRAM 30 m telescopes. 
Combining our new data with the data of H$_2$CO ($2_{02}-1_{01}$) and 
CS ($2-1$ and $3-2$) previously reported by \citet{Watanabe2014}, 
we find that the emitting regions of H$_2$CO and CS have 
the relatively low H$_2$ density of $(0.6-2.6)\times10^4$ cm$^{-3}$ 
and $(2.9-12)\times10^4$ cm$^{-3}$, respectively. This indicates that 
these two species are not concentrated in the star-forming cores, 
but are widely distributed over the GMCs. Models assuming the lognormal 
H$_2$ density distribution with fixed optical depths also support this picture. 
The different H$_2$ densities derived for H$_2$CO and CS imply 
their different distributions, and probably different formation processes. 

\acknowledgments

We are grateful to the anonymous referee for his/her comments and suggestions 
to improve the paper. We thank the staff of the NRO 45 m telescope for 
excellent support. This work is supported by NAOJ ALMA Scientific Research 
Grant Number 2017-06B and JSPS KAKENHI Grant Number JP18K13577. 

\facility{NRO 45 m}

\software{Newstar}

\end{document}